\documentclass[%
aip,
amsmath,amssymb,
reprint,%
]{revtex4-1}

\usepackage{graphicx}% Include figure files
\usepackage{epsfig, subfigure}
\usepackage{dcolumn}% Align table columns on decimal point
\usepackage{bm}% bold math
\usepackage[utf8]{inputenc}
\usepackage[T1]{fontenc}
\usepackage{mathptmx}
\usepackage{etoolbox}
\usepackage{amsthm,amsmath,amssymb}
\usepackage{mathrsfs}
\graphicspath{{figs/}}

\makeatletter
\def\@email#1#2{%
	\endgroup
	\patchcmd{\titleblock@produce}
	{\frontmatter@RRAPformat}
	{\frontmatter@RRAPformat{\produce@RRAP{*#1\href{mailto:#2}{#2}}}\frontmatter@RRAPformat}
	{}{}
}%
\makeatother
\begin{document}
	
	\preprint{AIP/123-QED}
	
	\title{Hypersonic curved compression ramp flows with bistable states}
	% Force line breaks with \\
	\author{Ming-Zhi Tang}		
	\affiliation{ 
		Hypervelocity Aerodynamics Institute, China Aerodynamics Research and Development Centre, Mianyang 621000, China.%\\This line break forced with \textbackslash\textbackslash
	}%	
	\author{Gang Wang }	
	\author{Yan-Chao Hu }%
	\altaffiliation[huyanchao@cardc.cn ]\\
	
	\author{Wen-Feng Zhou }
	\author{Zhu-xuan Xie }
	\affiliation{ 
		Hypervelocity Aerodynamics Institute, China Aerodynamics Research and Development Centre, Mianyang 621000, China.%\\This line break forced with \textbackslash\textbackslash
	}%	
	\author{Yan-Guang Yang }
	\altaffiliation[yangyanguang@cardc.cn ]\\	
	\affiliation{
		China Aerodynamics Research and Development Centre, Mianyang 621000, China.
	}%Lines break automatically or can be forced with \\
	
	\date{\today}% It is always \today, today,
	
	\begin{abstract}
		The bistable states and separation hysteresis in curved compression ramp (CCR) flows, and the corresponding aerothermal characteristics (including wall friction $C_{f}$, pressure and heat flux $St$), are studied numerically and theoretically. Direct numerical simulations of separation hysteresis induced by variation of turning angle $\phi$, as well as the influence of inflow Mach number and wall temperature on hysteresis loops, are carried out. Distributions of wall friction, pressure and heat flux are analyzed. Further, emergence of $C_{f}$’s  first and second minima in the separation bubble is interpreted, revealing it is dominated by the adverse pressure gradient (APG) induced by separation and reattachment shocks. The present results and analysis indicate that the reversed-flow singularity of Smith (Proceedings of the Royal Society of London. A. Mathematical and Physical Sciences, 1988, 420: 21-52) is less likely to occur in CCR flows. The prediction of peak pressure $p_{pk}$ of separation states confirms the model based on the minimum viscous dissipation theorem (Physics of Fluids, 2020, 32(10):101702). While the pressure overshoot $p_{os}$ can be analyzed by shock-polars with pressure match of compression and expansion process. The correlation between peak heat flux and peak pressure rise of both separation and attachment states is also discussed in terms of the classical power relations.

	\end{abstract}
	
	\maketitle
	
%	\begin{quotation}
%		The ``lead paragraph'' is encapsulated with the \LaTeX\ 
%		\verb+quotation+ environment and is formatted as a single paragraph before the first section heading. 
%		(The \verb+quotation+ environment reverts to its usual meaning after the first sectioning command.) 
%		Note that numbered references are allowed in the lead paragraph.
%		%
%		The lead paragraph will only be found in an article being prepared for the journal \textit{Chaos}.
%	\end{quotation}
	
	\section{\label{sec:intr}Introduction}
	
Deflection of control surface such as body flaps, elevons and rudders may cause intense shock wave/boundary layer interaction (SBLI). SBLI may yield significant flow separation and lead to significant decrease in control effectiveness and the excessive increase of the heat flux, and even lose control of aircraft\cite{simeonides1994experimental,simeonides1995experimental,babinsky2014shock,zhang2020hypersonic}. During this process, the flow hysteresis may be encountered, which can lead to multistable systems manifested as the dependence of a system on its evolutionary history, and ubiquitous in aerospace flow systems\cite{hu2020bistable}. Hysteresis can lead to multiple values of aerodynamic coefficients such as lift and drag in the hysteresis loop\cite{yang2008an,mccroskey1982unsteady,mueller1985the,biber1993hysteresis,mittal2000prediction}. The hysteresis of the regular and Mach reflection is found when changing the incident angle of the shock\cite{hornung1979transition,chpoun1995reconsideration,vuillon1995reconsideration,chpoun1995numerical,ivanov2001flow}, and the mechanism is explained by Hu et al. with minimal dissipation theory\cite{hu2021mechanism}. Recently, a separation/attachment hysteresis in curved compression ramp (CCR) flows induced by variation of attack angle is observed via numerical simulation\cite{hu2020bistable}. Mechanism of separation hysteresis in CCR flows is discussed by Zhou et al\cite{zhou2021mechanism}.\\\indent
As a classic configuration of SBLI, compression ramp flow has been much studied\cite{babinsky2014shock}. The free interaction theory \cite{chapman1958investigation,erdos1962shock} (FIT) established by Chapman et al. argues that the supersonic flow is influenced by the local boundary layer and the inviscid contiguous stream rather than the further development of the interaction. The triple deck theory formulated by Stewartson \& Williams\cite{stewartson1969self} and Neiland\cite{neiland1973asymptotic} provides the characteristic scale of each region and a series of equations to solve the exact physical quantities. By analytically deducing and numerically solving the triple deck equation, Smith\cite{smith1988a}, Smith \& Khorrami\cite{smith1991the} found that when the turning angle of the compression ramp is large enough, the separation bubble will become unstable and break down, and a singularity may arise within the reversed-flow region when the turning angle reaches a critical value. Korolev et al.\cite{korolev2002once}, Logue et al.\cite{logue2014instability} and Gai \& Khraibut\cite{gai2019hypersonic} confirmed the separation bubble will become unstable and break down, but they did not encounter singularity.\\\indent
The peak heat flux in the reattachment region is an important feature of separated flow. Most empirical methods to predict the peak heat flux always correlate it with peak pressure rise\cite{simeonides1994experimental,simeonides1995experimental,holden1978a,hung1973shockwave,hung1973interference}. Studies on wall pressure distributions are abundant. Theoretical predition mothed for pressure distribution of incipiently separated or just past incipient separation flows is proposed by Stollery \& Bate\cite{stollery1974turbulent}, which is not applicable for well separated flows. For well separated flow, FIT can predict the pressure rise at separation point and the plateau \cite{chapman1958investigation}. However, the peak pressure can not be predicted by FIT. Some semi-empirical expressions for plateau and peak pressure under certain conditions have been established\cite{gumand1959on}. Rencently, Hu et al. proposed an implicit model based on the minimum viscous dissipation (MVD) theorem for prediction of plateau and peak pressure of compression ramp flows with large separated regions\cite{hu2020prediction}. Li proposed an explicit theretical model combining FIT to predict plateau and peak pressure, which is validated at low-to-medium Reynolds number\cite{li2021theory}.\\\indent
Herein, separation hysteresis of CCR flows induced by the variation of turning angle $\phi$ is studied with two-dimensional (2D) numerical simulations. The hysteresis loops and the corresponding wall quantities are analyzed in the second section, as well as the influence of the inflow Mach number and wall temperature. In the third section, we discuss the aerothermal characteristics of the hysteresis process in detail.
	
	\section{\label{sec:res}Results}
	
	\subsection{\label{sec:hysloop}The hysteresis loops induced by turning angle $\phi$-variation}
		
\begin{figure*}[htbp] 
		\hspace{-10mm}
		\subfigure[\label{subfig:M6T15_hys_loop}]{\includegraphics[width = 0.72\columnwidth]{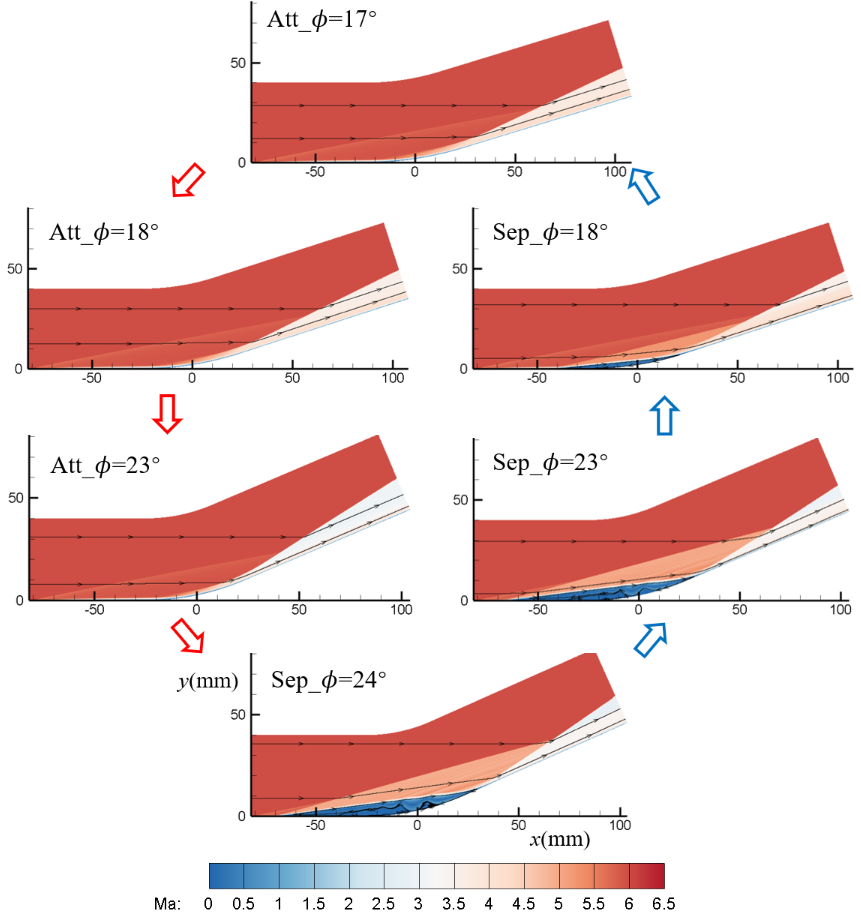}}
		\hspace{-1.5mm}		
		\subfigure[\label{subfig:M5T15_hys_loop}]{\includegraphics[width = 0.72\columnwidth]{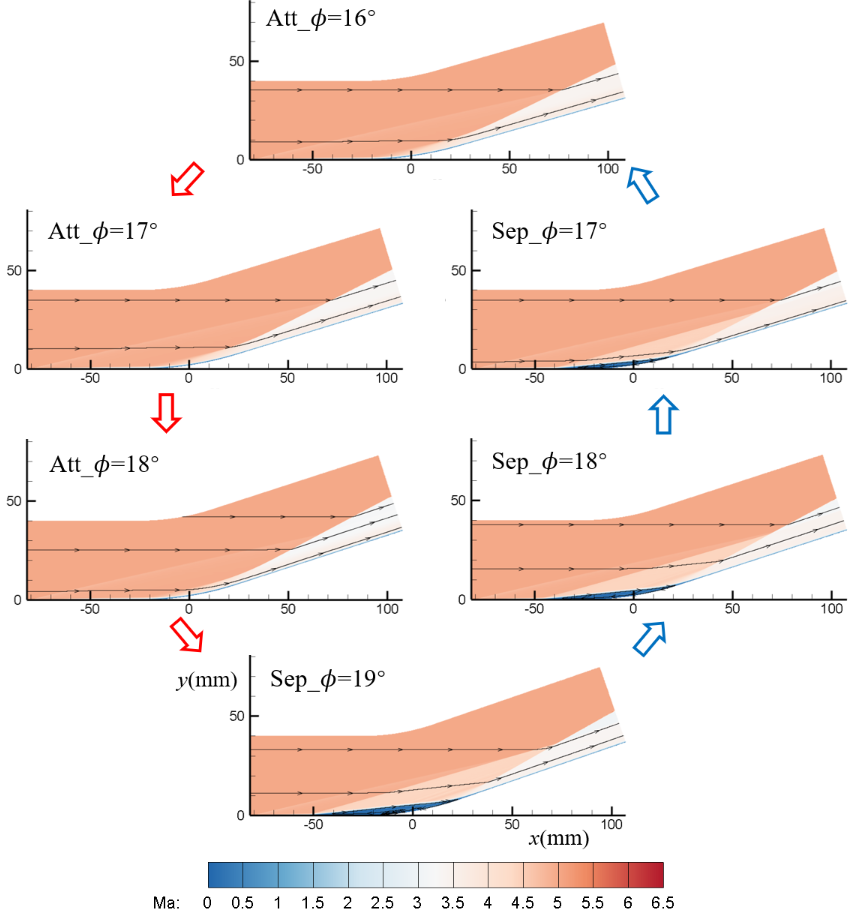}} 
		\hspace{-2mm}		
		\subfigure[\label{subfig:M6T2_hys_loop}]{\includegraphics[width = 0.72\columnwidth]{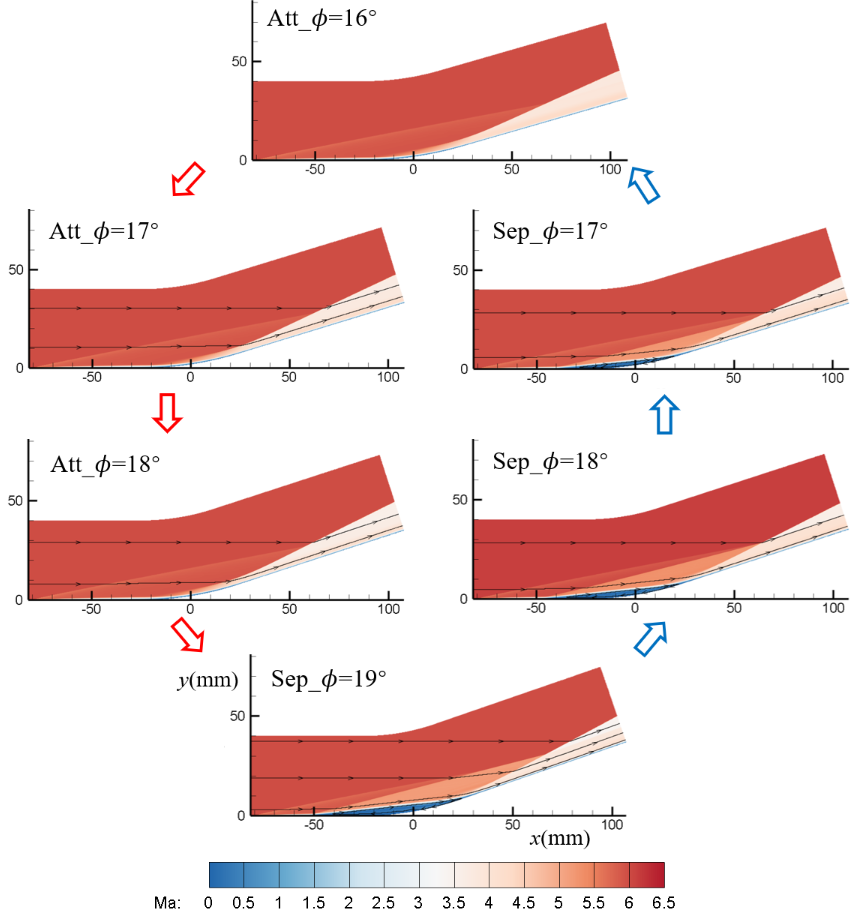}}
		\hspace{-10mm}
		\caption{Separation/attachment hysteresis loop induced by $\phi$ with different $Ma_{\infty}$ and $T_{w}$ conditions. The black lines with arrows are the streamlines. The blue regions are separation bubble. (a)$Ma_{\infty}=6.0,T_{w}=1.5$ (b)$Ma_{\infty}=5.0,T_{w}=1.5$ (c)$Ma_{\infty}=6.0,T_{w}=2.0$} \label{fig:hys_loop}
	\end{figure*}

\begin{figure*}[htbp] 
		\subfigure[\label{subfig:con_his}]{\includegraphics[width = 0.92\columnwidth]{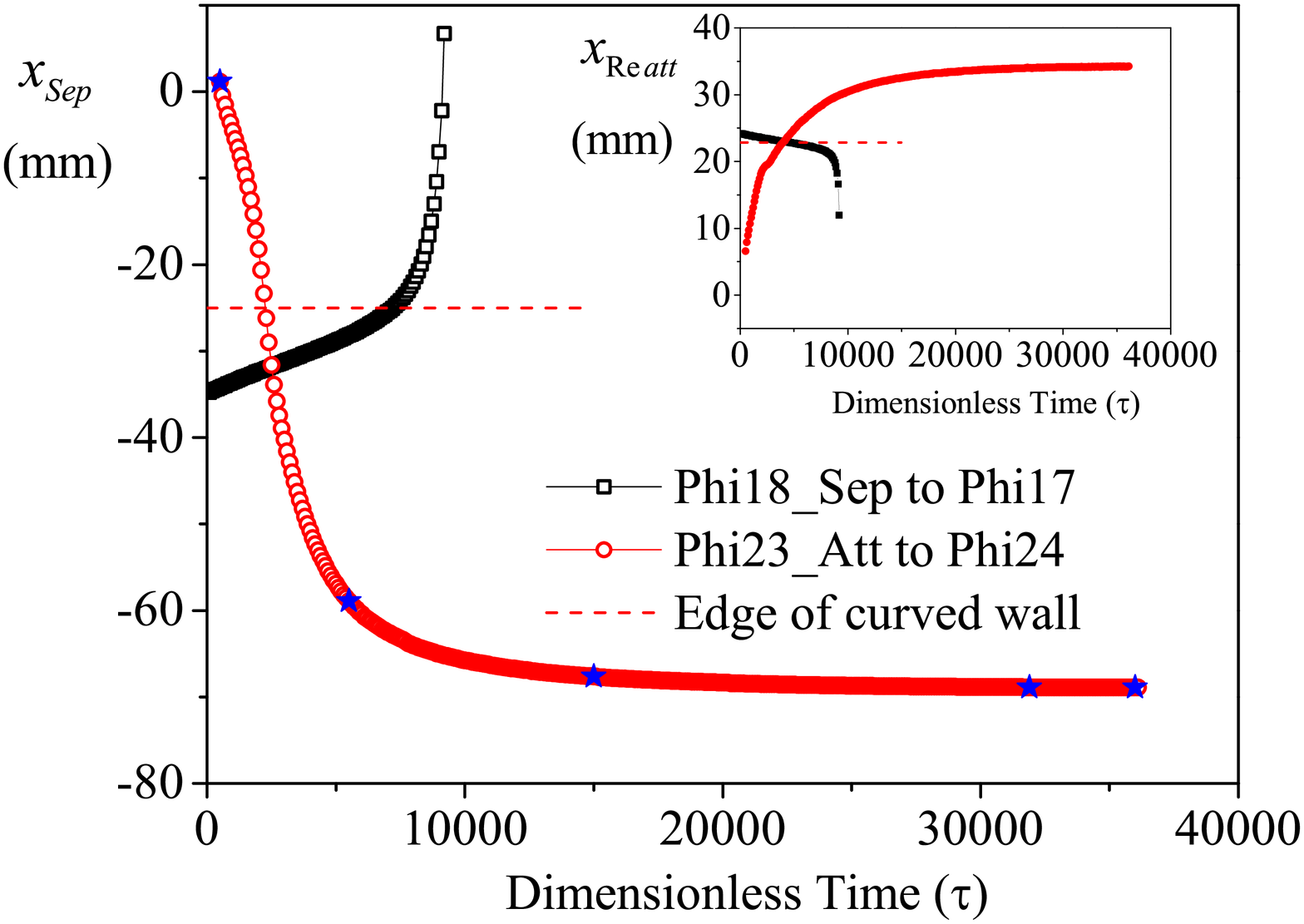}}
		%	\hspace{-2mm}
		\subfigure[\label{subfig:con_his_Ch}]{\includegraphics[width = 0.92\columnwidth]{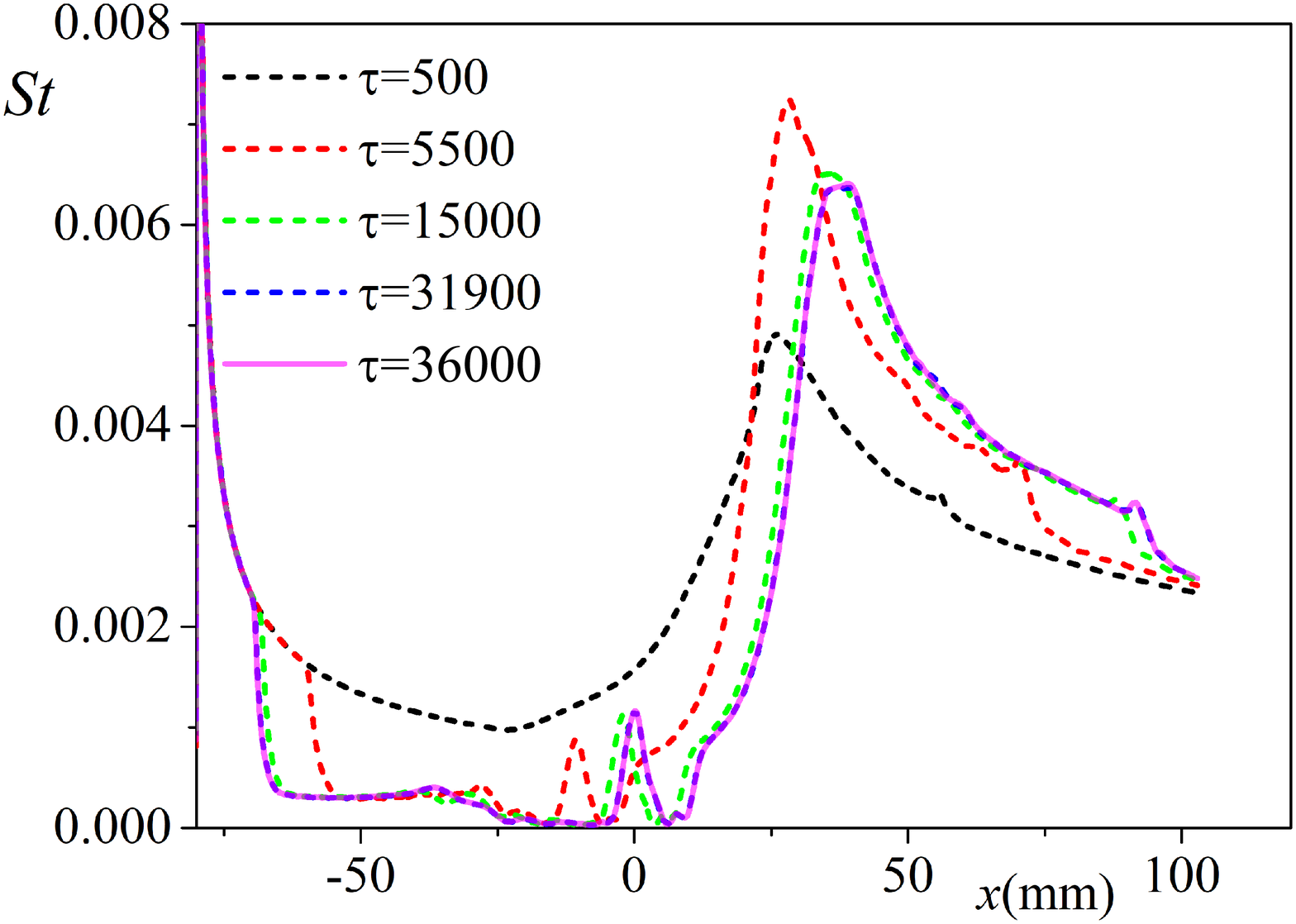}}
		%	\vspace{5mm}
		\caption{Time convergence process for $Ma_{\infty}=6.0,T_{w}=1.5$. (a) Separation/reattachment locations. The red circles are the separation/reattachment locations from $\phi = 23^{\circ}$ attachment state to $\phi = 24^{\circ}$. The black squares are the separation/reattachment locations from $\phi = 18^{\circ}$ separation state to $\phi = 17^{\circ}$. The solid and hollow symbols represent reattachment and separation location, respectively. (b) Convergence of the wall heat flux coefficient $C_{h}$ during the process from $\phi = 23^{\circ}$ attachment state to $\phi = 24^{\circ}$.} \label{fig:converg_his}
	\end{figure*}

	Direct numerical simulations (DNS) of 2D CCR flow based on OpenCFD\cite{li2010direct} are carried out in the present research. The flat plate starts at $x=-80mm$. The turning angle $\phi$ ranges from $16^{\circ}$ to $24^{\circ}$, and the curved wall is an arc with the curvature radius $R=\frac{L}{\sin\frac{\phi}{2}}$, where $L=25mm$ with $x=-L$ being the starting point of the curved wall. The unit Reynolds number $Re_{\infty}$ ($1 mm^{-1}$) of the flow is $3000$, $Pr = 0.7$, the specific heat ratio $\gamma = 1.4$, the inflow Mach number $Ma_{\infty}=6.0$, and the normalized wall temperature $T_{w}=1.5$ ($T_{w} = \hat{T}_{w}/\hat{T}_{0}$, where $\hat{T}_{0}=108.1K$ is the inflow temperature). Simulations are also carried out with inflow and boundary conditions (IBCs) of ($Ma_{\infty}=5.0,T_{w}=1.5$) and ($Ma_{\infty}=6.0,T_{w}=2.0$). Other flow settings are consistent with Ref.\cite{hu2020bistable} and numerical validations have been provided in Refs.\cite{hu2020bistable,li2010direct}.\\\indent
	Fig.\ref{subfig:M6T15_hys_loop} shows the separation/attachment hysteresis loop induced by $\phi$-variation, with IBCs of ($Ma_{\infty}=6.0,T_{w}=1.5$). A steady attachment flow is organized when $\phi = 17^{\circ}$. And the attachment state maintains till $\phi = 23^{\circ}$ in the process that $\phi$ increases with $\Delta \phi = 1^{\circ}$ gradually after the flow reaches convergence (state Att\_$\phi = 17^{\circ}$ to Att\_$\phi = 23^{\circ}$ in Fig.\ref{subfig:M6T15_hys_loop}). When $\phi$ reaches $24^{\circ}$, the flow will suddenly separate. When $\phi$ decreases from $24^{\circ}$ with the same $\Delta \phi$ after the flow reaches a steady state, the flow maintains separation till $\phi$ reaches $17^{\circ}$, at which angle the separation will disappear and the flow reaches attachment state again. When $\phi \in (\phi_{a},\phi_{s})$ ($\phi_{a} = 17^{\circ}$ and $\phi_{s} = 24^{\circ}$ for the current IBCs), the flow states may be either separated or attached with the same IBCs, depending on different initial conditions.\\\indent
	Locations of separation/reattachment points and distributions of Stanton number $St$ are monitored to determine
	convergence, and their values during the process from separation to attachment (Phi18\_Sep to Phi17) and from attachment to separation (Phi23\_Att to Phi24) are depicted in Fig.\ref{subfig:con_his} and Fig.\ref{subfig:con_his_Ch}. The Stanton number is
	\begin{equation} \label{eq:st}
		St = \frac{q_{w}}{\rho_{\infty} u_{\infty} (h_{\infty} + r u^{2}_{\infty} /2 - h_{w})}
	\end{equation}
	where $q_{w}$ is the wall heat flux, $r= \sqrt{Pr}$, $\rho_{\infty}$ and $u_{\infty}$ the inflow density and velocity. For the separation states, the flow is said to be in convergence when the displacements of separation point and the location of peak heat flux are less than $0.01mm$ in $1000 \tau$. (Dimensionless time $1 \tau$ corresponds to $1/(Ma_{\infty}\sqrt{\gamma R_{g} \hat{T}_{ref}}$)  , and the flow passes through $1mm$ in $1 \tau$.) An interesting phenomenon is that separation/reattachment points will not lie on the curved wall after convergence in separation states.\\\indent
	Similar hysteresis processes are observed with IBCs ($Ma_{\infty}=5.0,T_{w}=1.5$) and ($Ma_{\infty}=6.0,T_{w}=1.5$), as shown in Fig.\ref{subfig:M5T15_hys_loop} and Fig.\ref{subfig:M6T2_hys_loop}. But the interval $(\phi_{a},\phi_{s})$ is shifted and narrowed to $(16^{\circ},19^{\circ})$ for the two cases. Decrease of $Ma_{\infty}$ or increase of $T_{w}$ results in decrease of maximum APG the boundary layer can resist\cite{zhou2021mechanism}. Therefore separation will appear at a smaller $\phi$.\\\indent 
	The closed separation/attachment hysteresis loop is composed of three parameter intervals of turning angle $\phi$ with certain $Ma_{\infty}$, $Re_{\infty}$ and $T_{w}$, i.e., overall separation interval (OSI) in which the flow can only be in separation state; overall attachment interval (OAI) in which the flow can only be in attachment state; dual-solution interval (DSI) in which both steady attachment and separation states could exist. Decreasing $Ma_{\infty}$ or increasing $T_{w}$ will shift and reduce the DSI $(\phi_{a},\phi_{s})$.\\

	\subsection{Distributions of $C_{f}$, $p_{w}/p_{\infty}$ and $St$}

\begin{figure*}[htbp]
		\subfigure[\label{subfig:M6Tw1.5_cf}]{\includegraphics[width = 0.65\columnwidth]{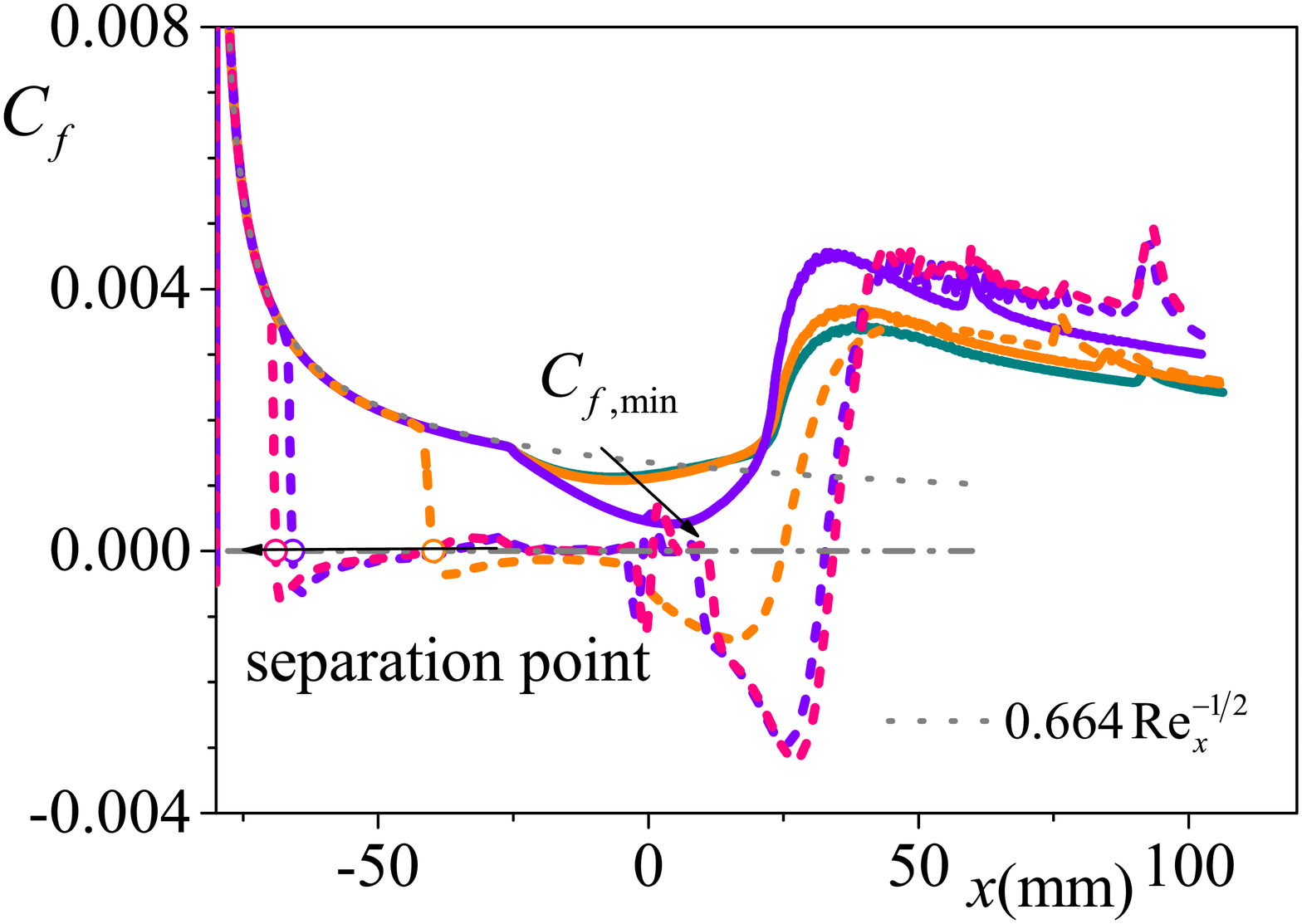}}
		\hspace{0em}
		\subfigure[\label{subfig:M6Tw1.5_Pw}]{\includegraphics[width = 0.65\columnwidth]{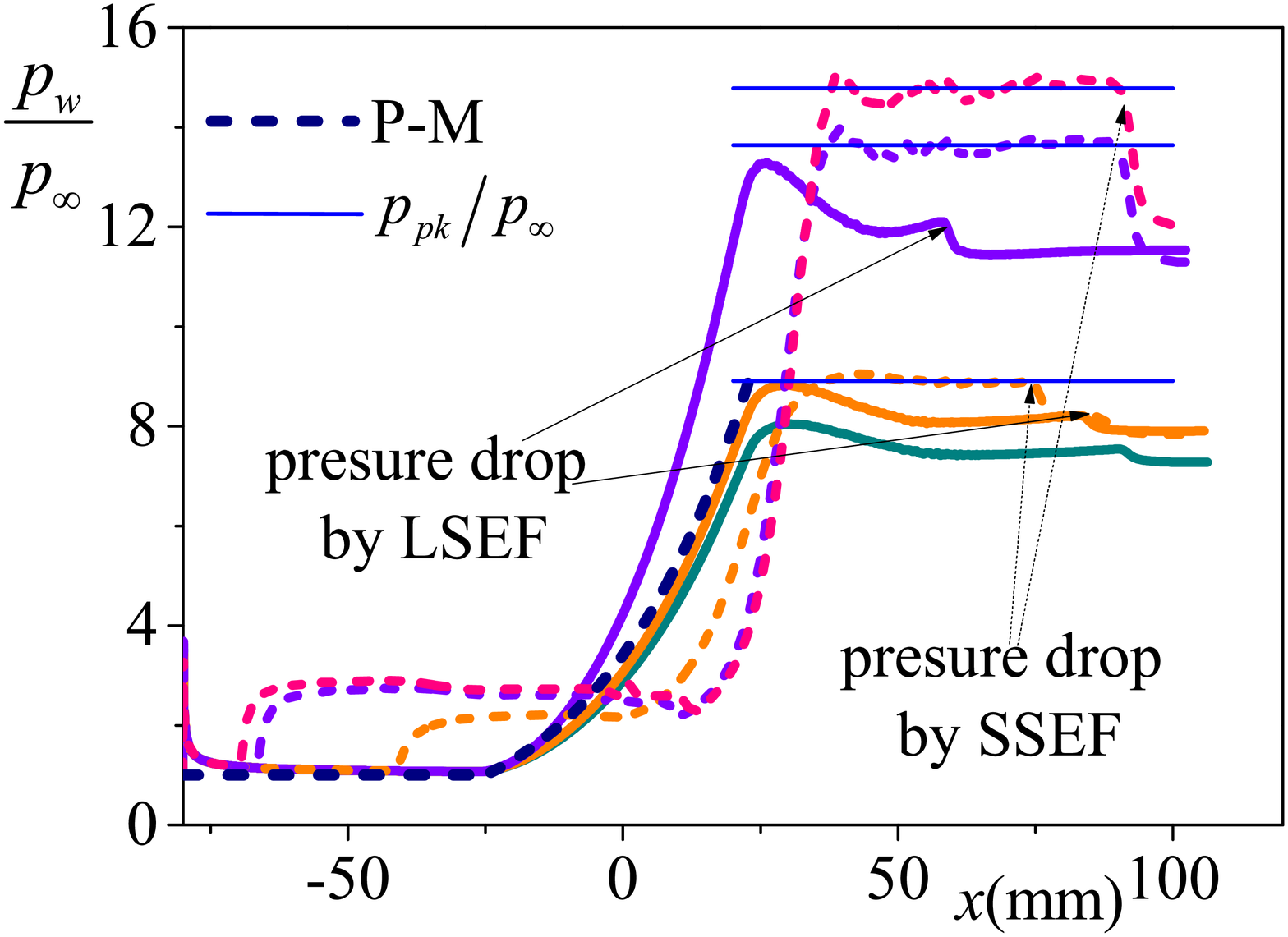}}
		\hspace{0em}
		\subfigure[\label{subfig:M6Tw1.5_qw}]{\includegraphics[width = 0.65\columnwidth]{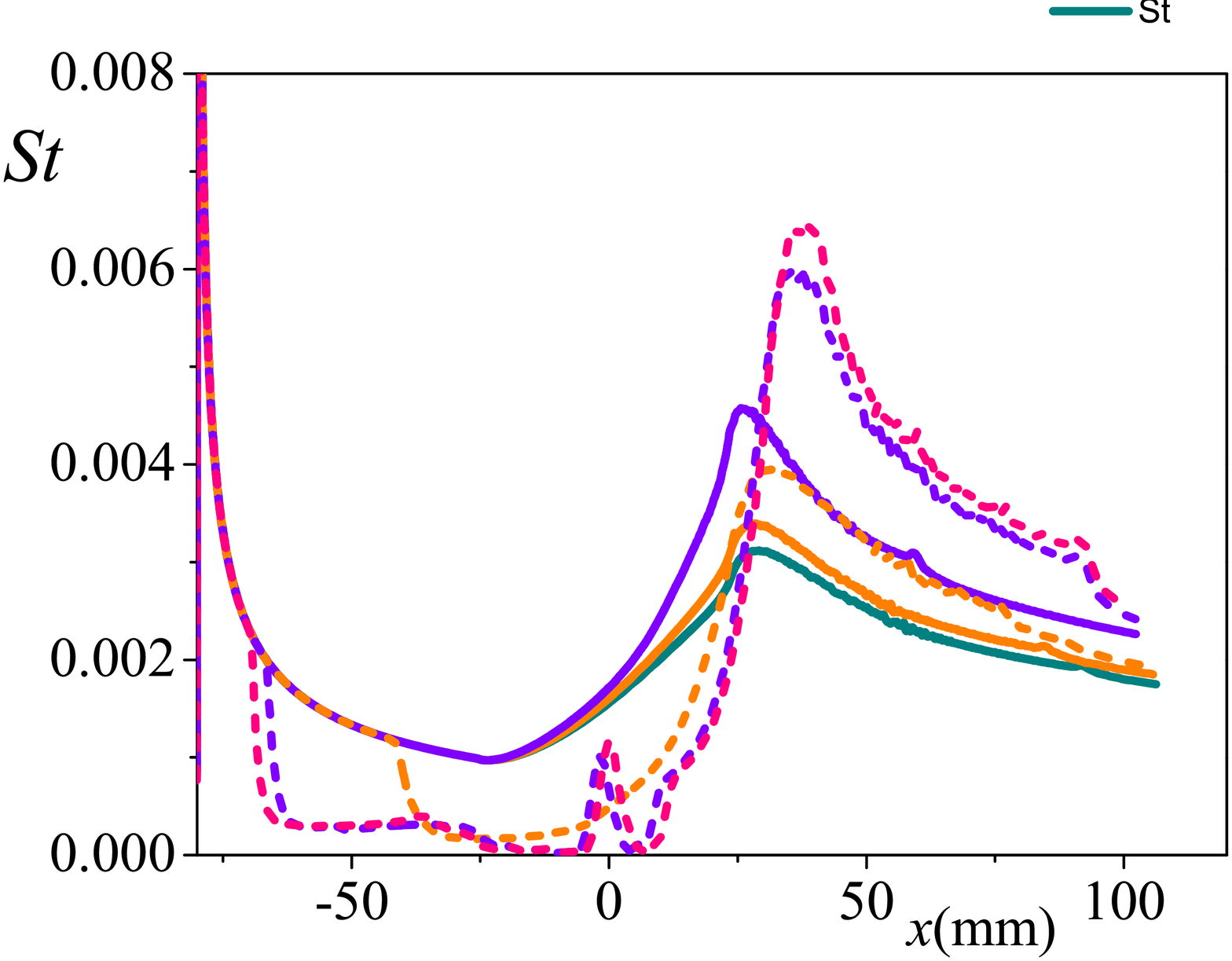}}
		
		\subfigure[\label{subfig:M5Tw1.5_cf}]{\includegraphics[width = 0.65\columnwidth]{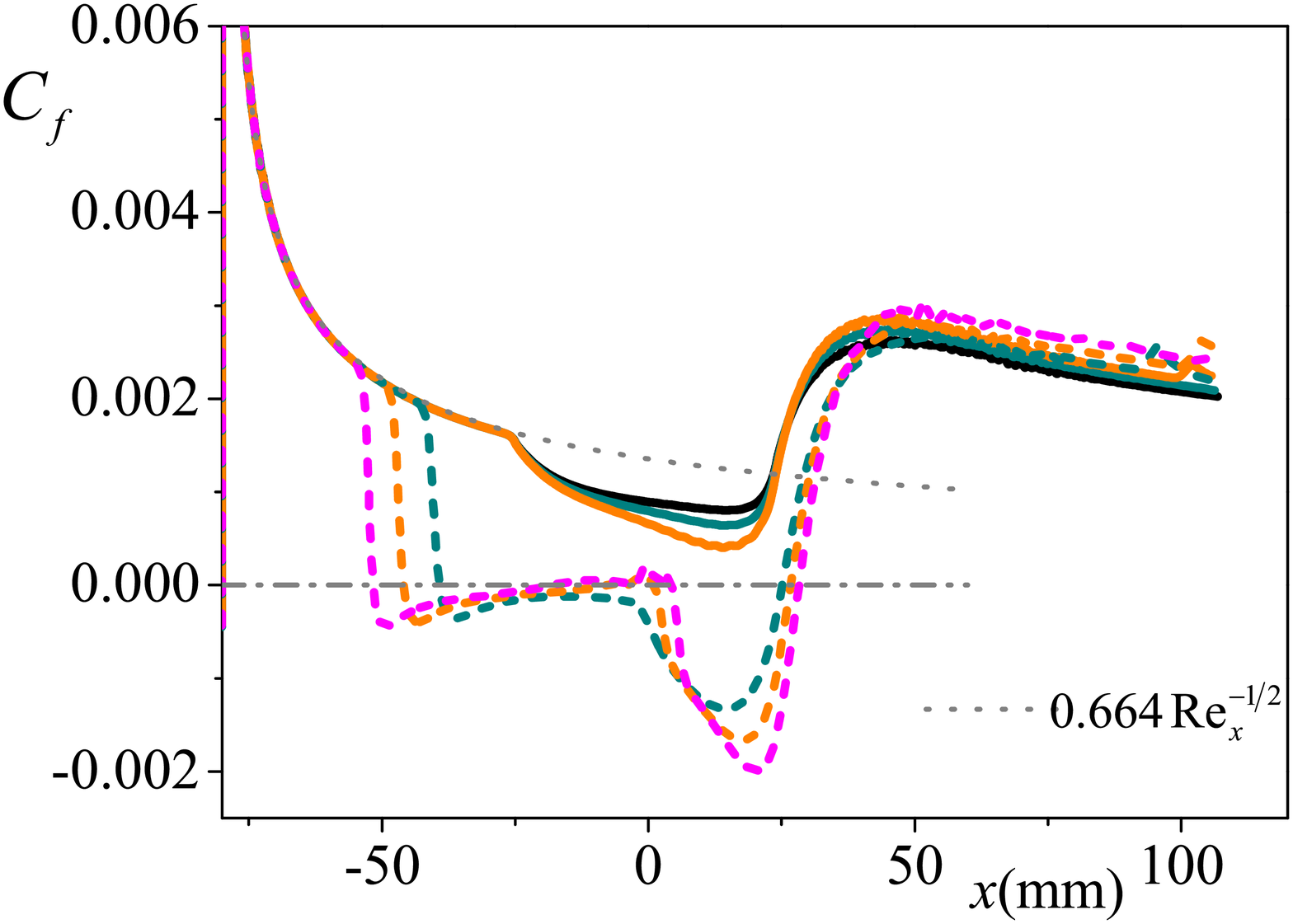}}
		\hspace{0em}
		\subfigure[\label{subfig:M5Tw1.5_Pw}]{\includegraphics[width = 0.65\columnwidth]{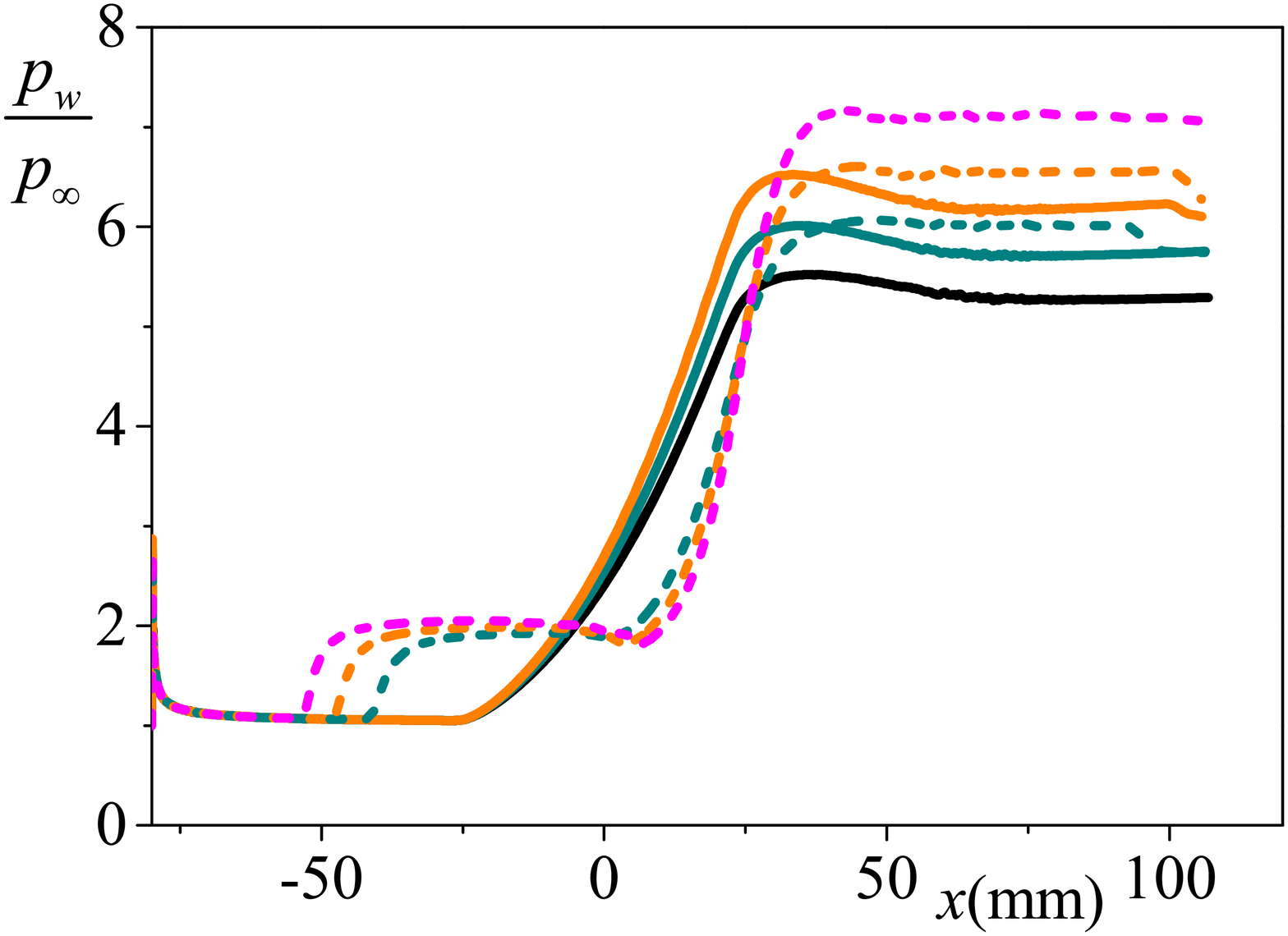}}
		\hspace{0em}
		\subfigure[\label{subfig:M5Tw1.5_qw}]{\includegraphics[width = 0.65\columnwidth]{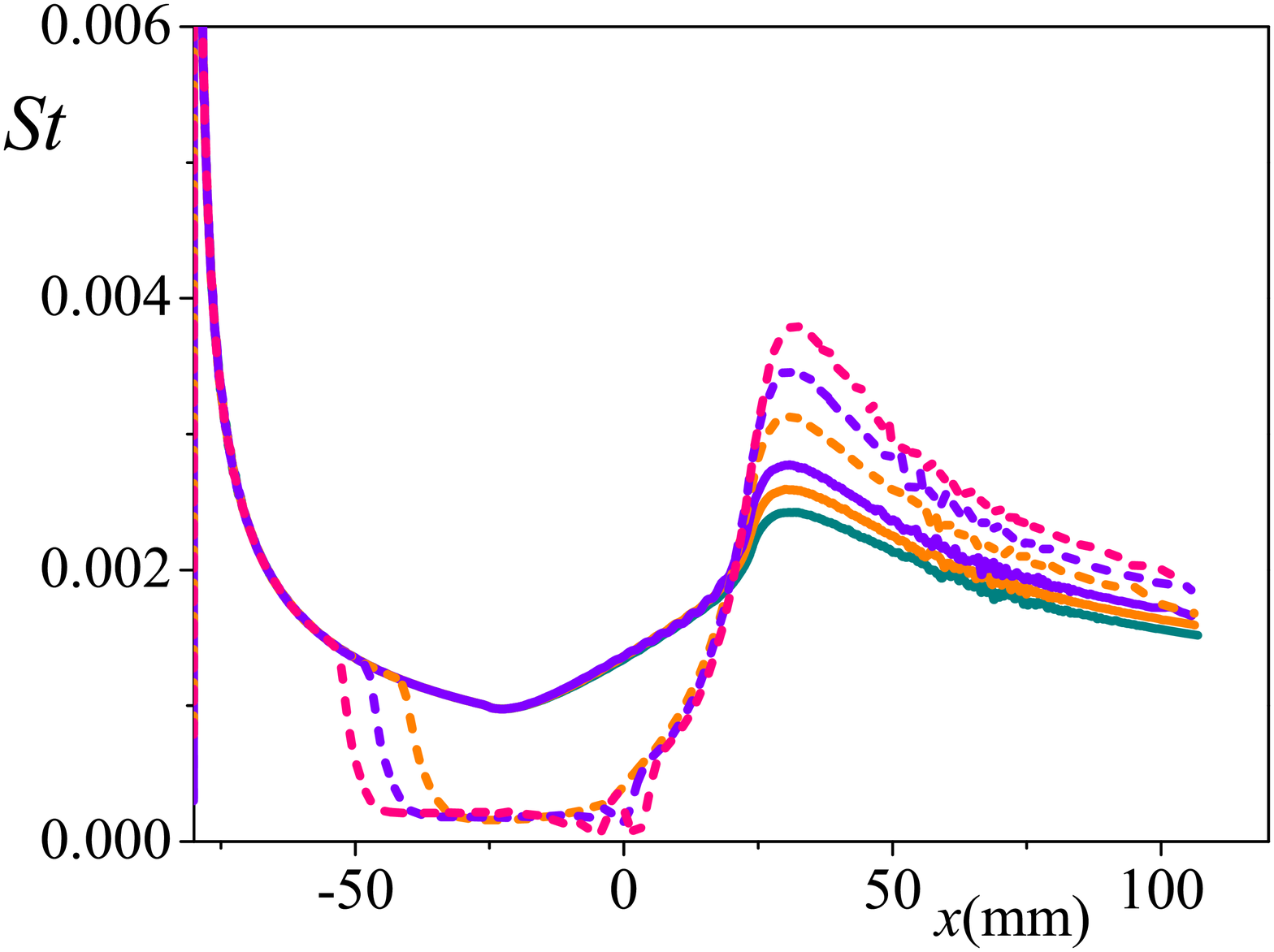}}
		
		\subfigure[\label{subfig:M6Tw2.0_cf}]{\includegraphics[width = 0.65\columnwidth]{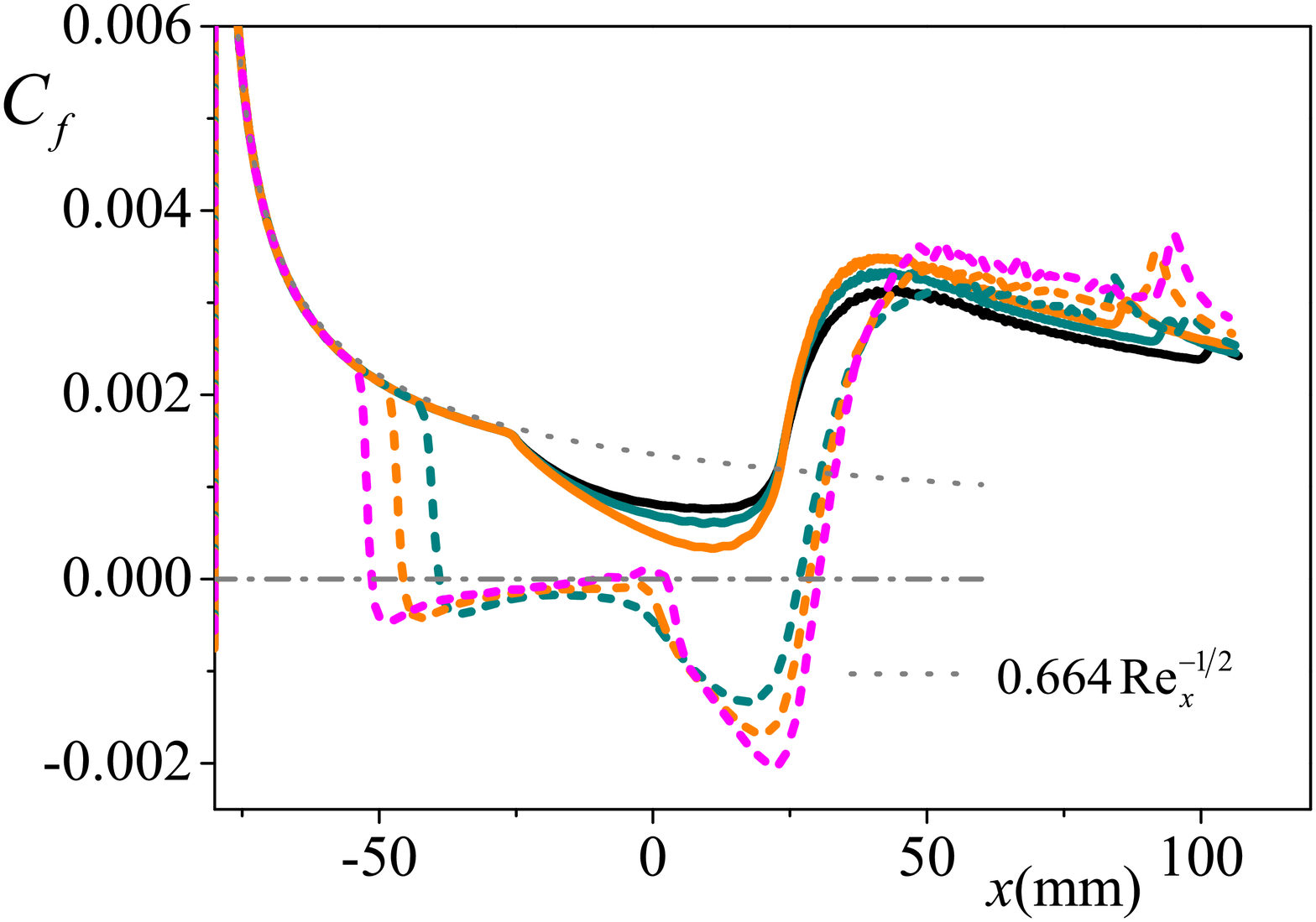}}
		\hspace{0em}
		\subfigure[\label{subfig:M6Tw2.0_Pw}]{\includegraphics[width = 0.65\columnwidth]{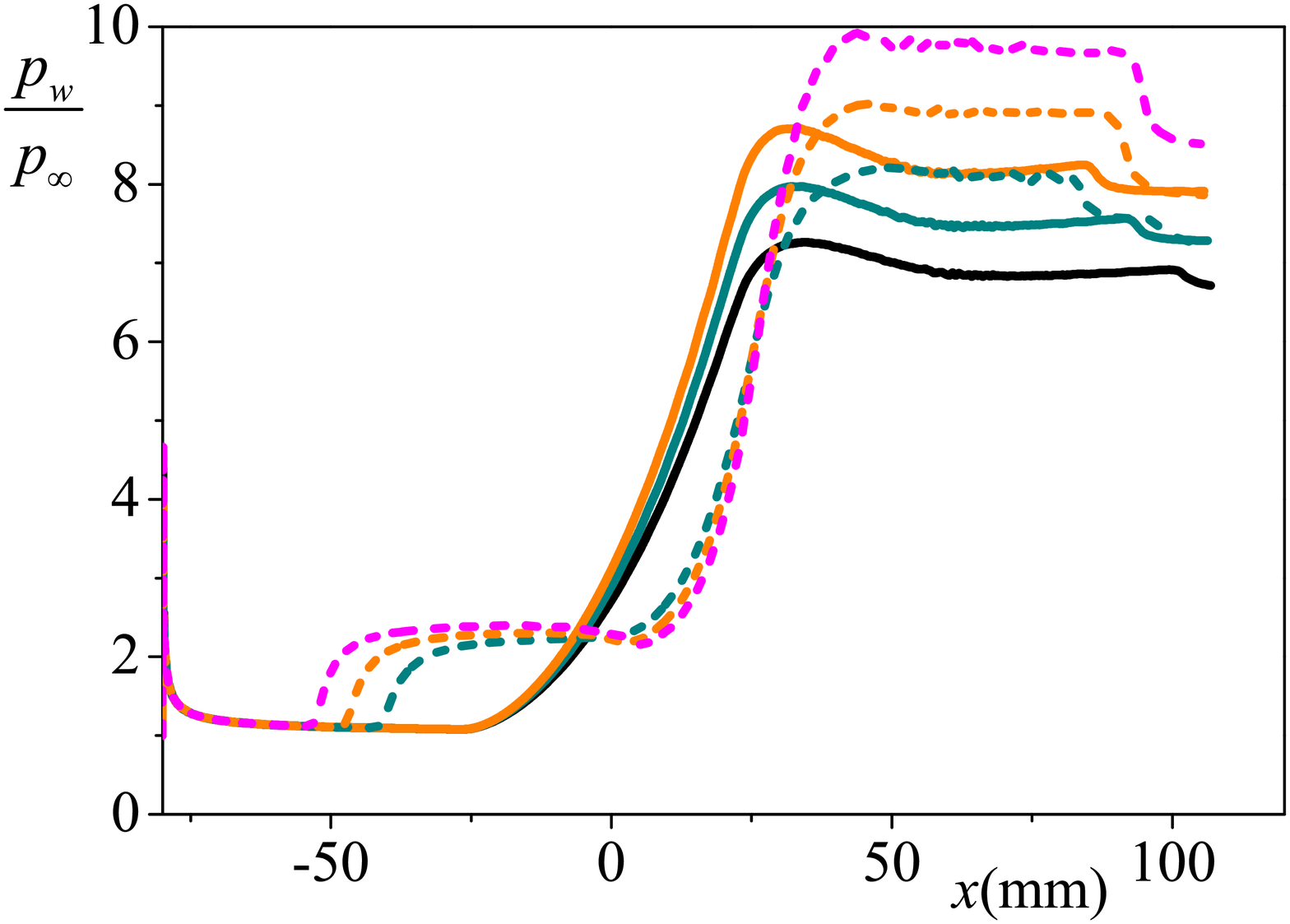}}
		\hspace{0em}
		\subfigure[\label{subfig:M6Tw2.0_qw}]{\includegraphics[width = 0.65\columnwidth]{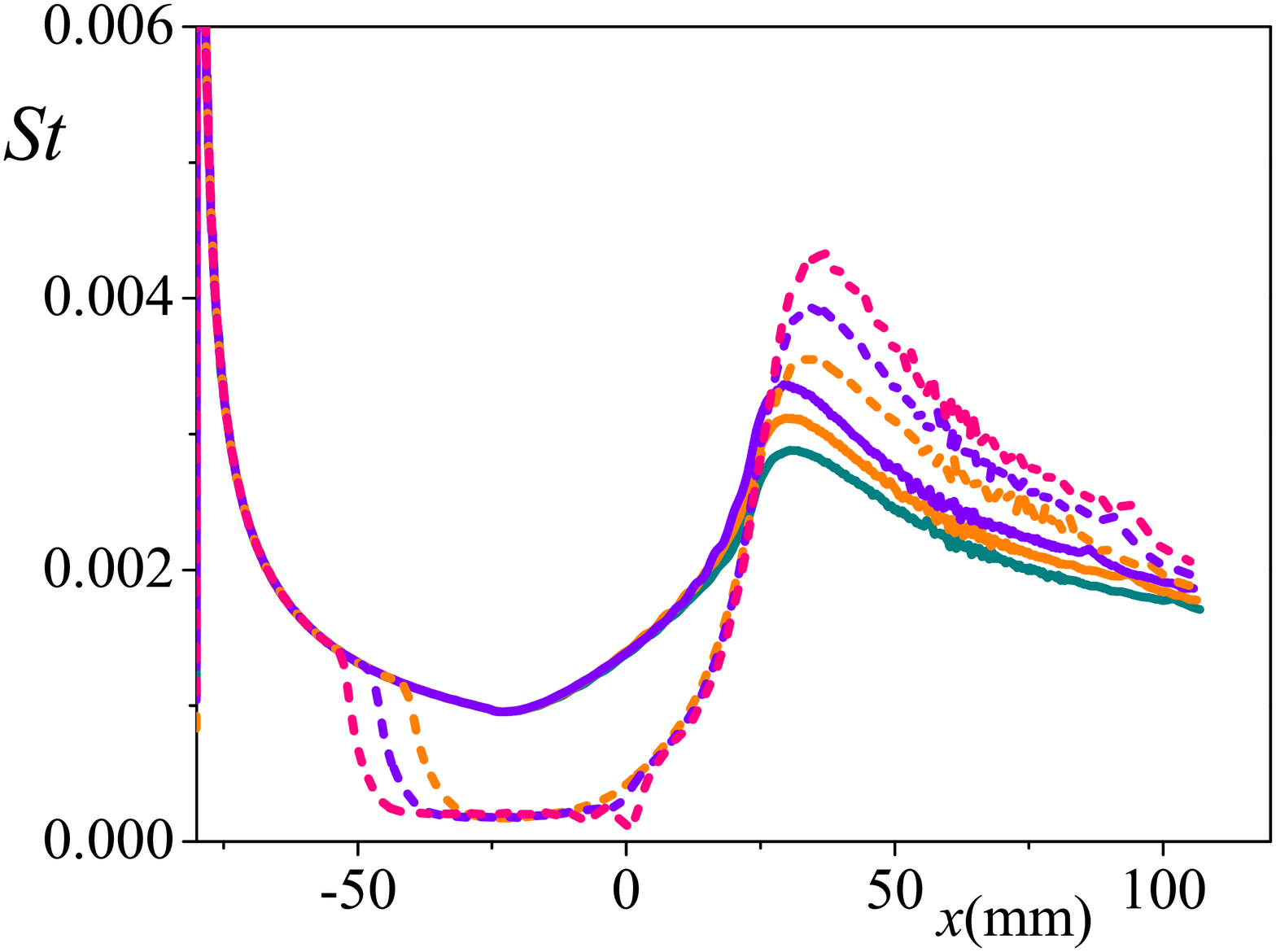}}	
		\caption{Distributions of wall friction coefficient $C_{f}$, normalized wall pressure $p_{w}/p_{\infty}$, Stanton number $St$ for attachment and separation states of different turning angle $/phi$. Upper panel: $Ma_{\infty}=6.0,T_{w}=1.5$, (a) $C_{f}$; (b )$p_{w}/p_{\infty}$; (c) $St$. middle panel: $Ma_{\infty}=5.0,T_{w}=1.5$, (d) $C_{f}$; (e) $p_{w}/p_{\infty}$; (f) $St$. low panel: $Ma_{\infty}=6.0,T_{w}=2.0$, (g) $C_{f}$; (h) $p_{w}/p_{\infty}$; (i) $St$. Black: $16^{\circ}$; dark cyan: $17^{\circ}$; orange: $18^{\circ}$; magenta: $19^{\circ}$; violet: $23^{\circ}$;pink: $24^{\circ}$. Solid lines for attachment states, and dash lines for separation states.} \label{fig:CfPwQ}
	\end{figure*}

	Distributions of wall friction $C_{f}$ with different $\phi$ are shown in Fig.\ref{subfig:M6Tw1.5_cf}, Fig.\ref{subfig:M5Tw1.5_cf} and  Fig.\ref{subfig:M6Tw2.0_cf} for different $Ma_{\infty}$ and $T_{w}$ conditions, where $C_{f}=\frac{2\tau_{w}}{\rho_{\infty} u_{\infty}^2}$ and $\tau_{w}=\mu_{w} \frac{\partial u}{\partial y}|_{w}$ is the wall shear stress. For attachment states, the non-monotonic distributions of $C_{f}$ on the curved wall result from the compression and APG effects simultaneously. The minimum of $C_{f}$ decreases with the increase of $\phi$, indicating a tendency to separate. For separation states, the separation point moves upstream with increasing $\phi$, to obtain higher wall friction to balance APG, corresponding to a larger separation bubble. Due to the breakdown of vortices, fluctuations of $C_{f}$ arise near $x=0$ for $\phi = 23^{\circ}$ and $24^{\circ}$. In the downstream of curved wall, $C_{f}$ reaches maximum, and then decreases with streamwise development in both separation and attachment states.
	\par The wall pressure $p_{w}/p_{\infty}$  distributions are shown in Fig.\ref{subfig:M6Tw1.5_Pw}, Fig.\ref{subfig:M5Tw1.5_Pw} and Fig.\ref{subfig:M6Tw2.0_Pw}, where $p_{\infty}$ is the inflow pressure. The attachment states sustain approximate isentropic compression process induced by the curve wall. As shown by the navy dash line in Fig.\ref{subfig:M6Tw1.5_Pw}, the process can be described by Prandtl-Meyer relation
	\begin{equation} \label{eq:PM}
		\begin{array}{c}
			\nu [Ma(\varphi)] - \nu (Ma_{\infty}) = -\varphi \\
			\frac{p_{w}(\varphi)}{p_{\infty}} = \left\{ \frac{\vartheta(Ma_{\infty})}{\vartheta \left[ Ma(\varphi) \right]} \right\}^{\frac{\gamma}{\gamma -1}} \\
			\vartheta(Ma) = 1 + \frac{\gamma -1}{2}Ma^2
		\end{array}
	\end{equation}
	where $\varphi$ is the turning angle at $x$  
	\begin{equation}
		\varphi = \arcsin \left( \frac{x+L}{R} \right) \in [0, \phi]  \label{eq:phi}
	\end{equation}
	In the separation states, a pressure plateau  arises at the separation region and sustain a second pressure rise to reach the pressure peak $p_{pk}$ after reattaching.
	\par In the downstream of the curved wall, $p_{w}/p_{\infty}$  drops twice in attachment states or separation states with small $\phi$. The resultant favourable pressure gradient results in local maximum $C_{f}$. The second drop is due to the expansion fan induced by the leading edge (LE) shock (LESF), but the first drop results from different mechanism. For the attachment states, it is due to the pressure match of isentropic compression and pressure rise by the shock, which leads to an overshoot and will be discussed in the following. For the separation states, the expansion fan induced by separation shock (SSEF) is the major reason. Separation states with large $\phi$ encounter one pressure drop due to SSEF. After these processes, $p_{w}/p_{\infty}$  reaches the inviscid pressure rise.
	\par The Stanton number $St$ distributions are shown in Fig.\ref{subfig:M6Tw1.5_qw}, Fig.\ref{subfig:M5Tw1.5_qw} and Fig.\ref{subfig:M6Tw2.0_qw}. In attachment states, the heat fluxes increase along the curve wall and reach the peak values near its end. In separation states, the heat flux is close to zero (the fluctuations near $x= 0$ are due to the breakdown of vortices) inside the separation bubble, and the peak heat fluxes occur nearby the boundary layer reattaching. Compared with the separation states, the peak heat fluxes of the attachment states are lower with the same IBCs and  $\phi$, and the discrepancy increases with larger $\phi$ (up to 32\% in present cases).\\\indent
	As shown in the distributions of $C_{f}$, $p_{w}/p_{\infty}$ and $St$, the attachment and separation states with the same IBCs correspond to vastly different aerothermal manifestation.

	\section{Discussion } \label{sec:dis}
	
	\subsection{Wall shear stress in the reversed-flow region} \label{subsec:wall shear stress}
\begin{figure}[htbp]
		\centering
		\includegraphics[width = 1.0\columnwidth]{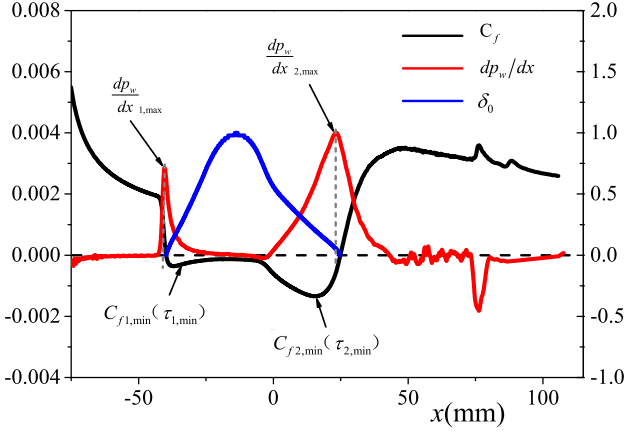} %\hspace{-2mm}
		\hspace{0mm}
		\caption{Distributions of wall friction (black line), nondimensionalized pressure gradient (red line) and nondimensionalized normal height of ZSSL (blue line) ($Ma_{\infty}=6.0,T_{w}=1.5,\phi=18^\circ$).} \label{fig:cf-dpdx}
	\end{figure}

	The appearance of the second minimum shear stress $\tau_{2,min}$ immediately before reattachment is a characteristic feature of compression ramp flows with large separation (CRFLS). Such phenomenon also exists in CCR flows with separation, as shown in Fig.\ref{fig:cf-dpdx}. As a measure
	of the extent of separation, the scaled angle $\alpha$ is defined\cite{stewartson1970on} as $\alpha = \frac{\phi Re^{1/4}}{C^{1/4} \lambda^{1/2} (Ma^{2}_{\infty}-1)^{1/2}}$, where $\phi$ is the geometric turning angle. $Re$ is the characteristic Reynolds number based on the flat plate length and the free-stream conditions. $C$ is the Chapman–Rubesin parameter and $\lambda = 0.332$ is the shear constant of a Blasius boundary layer.
	
\begin{figure}[htbp]
		\centering
		\includegraphics[width = 0.95\columnwidth]{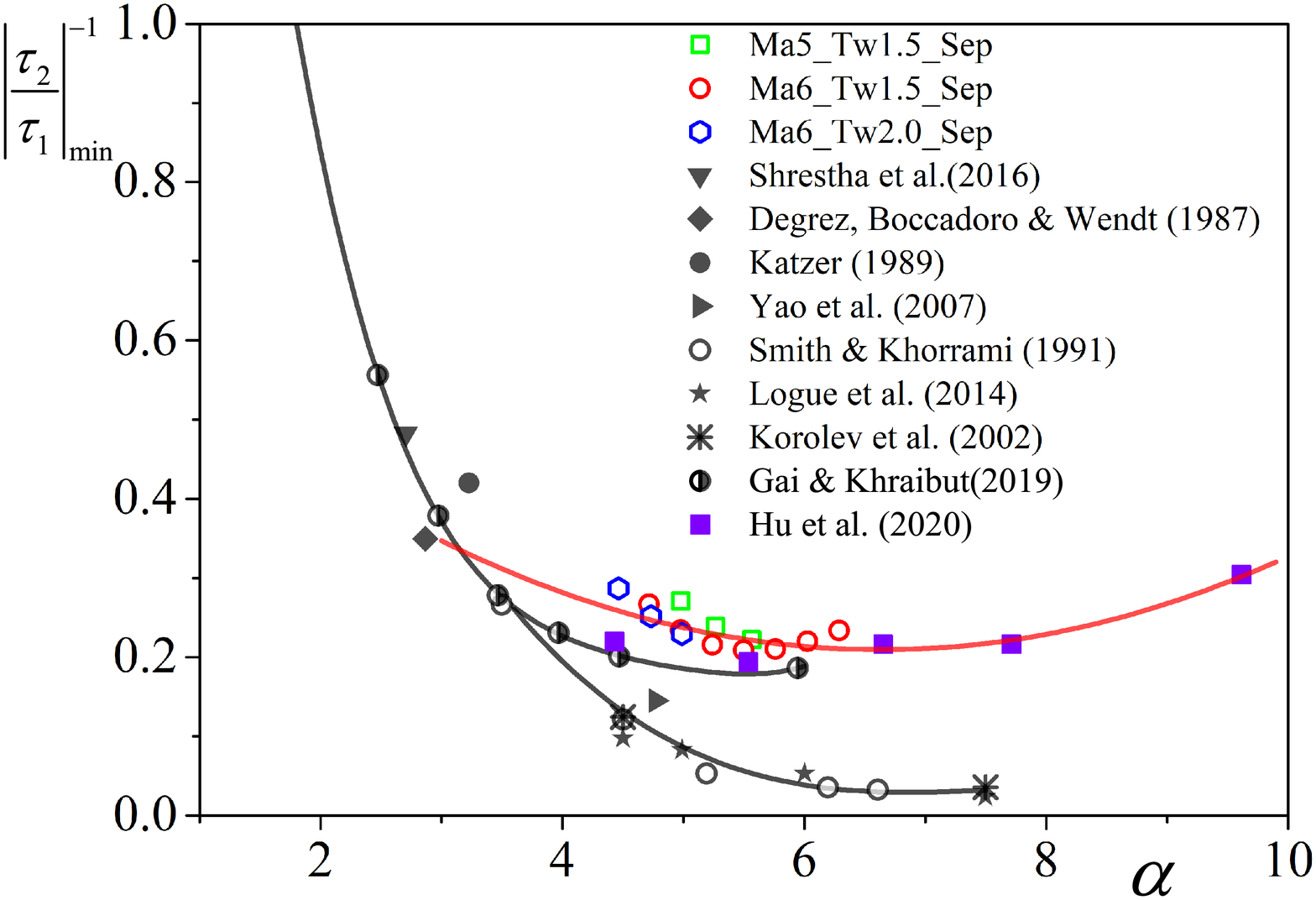} %\hspace{-2mm}
		\hspace{0mm}
		\caption{Variation of $|\tau_{2}⁄\tau_{1} |^{-1}_{min}$ with scaled angle $\alpha$.} \label{fig:alfa-tau_min}
	\end{figure}

	Smith\cite{smith1988a} and Smith \& Khorrami\cite{smith1991the} suggested that $\tau_{2,min}$ results from unstableness or breakdown of the separation bubble. Katzer\cite{katzer1989on} suggested it is the displacement of the separation bubble center that generates $\tau_{2,min}$. But they did not offer correlations of $\tau_{2,min}$ with other flow parameters. Smith\cite{smith1988a} and Smith \& Khorrami\cite{smith1991the} also suggested that when $\alpha$ is large enough, a singularity may arise with $|\tau_{2,min} | \to \infty$. Based on Neiland’s reattachment theory\cite{neiland1973asymptotic}, Korolev et al.\cite{korolev2002once} verified $|\tau_{2,min} | \sim \alpha$ when $\alpha \to \infty$, but they did not encounter singular for $\alpha$ up to 7.5. Gai \& Khraibut\cite{gai2019hypersonic} suggested that $|\tau_{2,min} | \to \infty$ is less likely to occur for low wall temperature ratio. Fig.\ref{fig:alfa-tau_min} shows the distributions of $|\tau_{2}⁄\tau_{1} |^{-1}_{min}$ (including the early literature\cite{smits1996turbulent,korolev2002once,logue2014instability,gai2019hypersonic,hu2020prediction,shrestha2016interaction,degrez1987the,katzer1989on,yao2007the}) with variation of $\alpha$. For the present cases, distribution of $|\tau_{2}⁄\tau_{1} |^{-1}_{min}$ is similar with the results of Gai \& Khraibut\cite{gai2019hypersonic} and Hu et al.\cite{hu2020prediction}. And $|\tau_{2}⁄\tau_{1} |^{-1}_{min}$  seems to increase slowly when $\alpha>6$ in the present cases and the results Hu et al.\cite{hu2020prediction}.
	
\begin{figure}[htbp]
		\centering
		\includegraphics[width = 0.95\columnwidth]{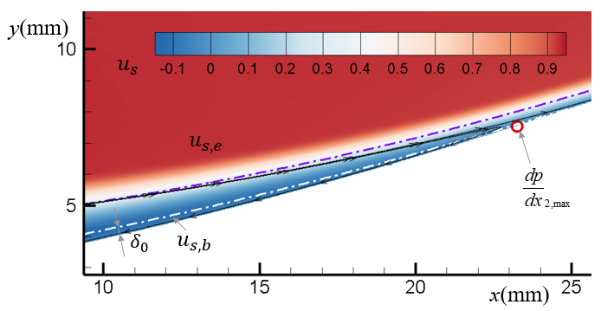} %\hspace{-2mm}
		\hspace{0mm}
		\caption{Tangential velocity field and the definition of $\delta_{0}$. The purple dash-dot line is the sonic line. The white dash-dot line is the ZSSL. ($Ma_{\infty}=6.0,T_{w}=1.5,\phi=18^\circ$)} \label{fig:vel_delta0}
	\end{figure}

	In CCR flows, $\tau_{2,min}$ can appear without separation bubble breakdown. Here, a theoretical interpretation of $\tau_{1,min}$ and $\tau_{2,min}$ is proposed as follows. The tangential momentum equation in orthogonal curvilinear coordinates\cite{2017Boundary} is incorporated as
	\begin{equation}\label{eq:moment}
		\rho [\frac{u_{s}}{h} \frac{\partial u_{s}}{\partial s}+u_{n} \frac{\partial u_{s}}{\partial n} +\frac{\kappa u_{s} u_{n}}{h}]=-\frac{1}{h} \frac{dp}{dx} + \frac{\partial \tau}{\partial n}
	\end{equation}
	where $h=1+\kappa(s)n$ and $\kappa(s)$ is the wall surface curvature. We carry out the analysis in the region between the wall and zero shear stress line (ZSSL, white dash-dot line in Fig.\ref{fig:vel_delta0}), where the shear stress $\tau$ is equal to 0 in the separation bubble. As shown in Fig.\ref{fig:vel_delta0}, the velocity $u_{s,b}$  between the white dash-dot line and the wall is small compared with the eternal flow velocity $u_{s,e}$, fulfilling $u_{s,b} / u_{s,e} = O(10^{-1})$. Then, Eq.\ref{eq:moment} to describe the region we focus on can be simplified as
	\begin{equation}\label{eq:dp_tau} 
		\frac{1}{h} \frac{dp}{dx} = \frac{\partial \tau}{\partial n}
	\end{equation}
	Integrating Eq.\ref{eq:dp_tau} from the wall to ZSSL along the normal direction $n$, with the assumption that the pressure is constant across the boundary layer, we can obtain
	\begin{equation} \label{eq:dp_del_tau}
		\frac{\ln[1+\kappa(s) \delta_{0}]}{\kappa(s)} \frac{dp}{dx} = \tau_{0} - \tau_{w}
	\end{equation}
	In the formula, the characteristic thickness $\delta_{0}$ corresponds to the normal distance from the wall to ZSSL, and $\tau_{0}=0$ is the shear stress on ZSSL. Besides, $\kappa(s) \delta_{0} \sim O(10^{-2})$, then $\ln[1+\kappa(s) \delta_{0}] \approx \kappa(s) \delta_{0}$. Combining the definition of $C_{f}$, Eq.\ref{eq:dp_del_tau} can be deduced and rearranged as
	\begin{equation} \label{eq:cf_dp_delta0}
		C_{f} \approx -\frac{2}{\rho_{\infty} u_{\infty}^2} \frac{dp_{w}}{dx} \delta_{0}
	\end{equation}
	\par
\begin{figure}[htbp]
		\centering
		\includegraphics[width = 0.95\columnwidth]{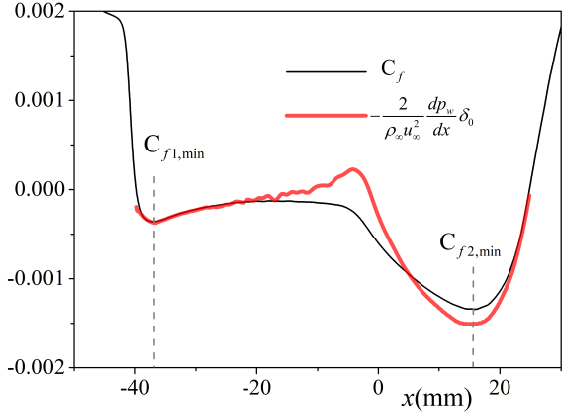} %\hspace{-2mm}
		\hspace{0mm}
		\caption{Comparison of the right-hand-side of Eq.\ref{eq:cf_dp_delta0} (red line) with $C_{f}$ (black line) from DNS results. ($Ma_{\infty}=6.0,T_{w}=1.5,\phi=18^\circ$)} \label{fig:cmp-cf-dpdx}
	\end{figure}

 From Eq.\ref{eq:cf_dp_delta0}, the wall friction is proportional to the product of APG and $\delta_{0}$. From Fig.\ref{fig:cf-dpdx} we can see that two local maximal APGs arise near separation and reattachment points, which are induced by separation and reattachment shocks, respectively. While $\delta_{0}$ is equal to 0 at separation and reattachment points, and have a maximum in the separation bubble. The location of maximum $\delta_{0}$ does not lie in the middle of separation and reattachment points(Fig.\ref{fig:cf-dpdx}), nor does $\delta_{0}$ present symmetric distribution, both manifesting the displacement of the separation bubble, as observed by Katzer\cite{katzer1989on}. The product of APG and $\delta_{0}$ results in two peak values, corresponding to the first and second minimum values of wall friction, i.e. $C_{f1,min}$ and $C_{f2,min}$. Therefore, the essential condition for the emergence of $C_{f2,min}$ may not be the unstableness or breakdown of separation bubble, but the explicit separation and reattachment shocks, which can also observed in the present case. Validation of Eq.\ref{eq:cf_dp_delta0} is depicted in Fig.\ref{fig:cmp-cf-dpdx}. Both the values and locations of $C_{f1,min}$ and $C_{f2,min}$ fit well, and the discrepancy of value is within 13\%. Moreover, as $\delta_{0} \to 0$ when approaching separation/reattachement point, the simplification from Eq.\ref{eq:moment} to Eq.\ref{eq:dp_tau} is closer to the real conditions, and better agreements can be obtained.
	\par From the perspective of the flow field, the sonic line of the separation state (purple dash-dot line in Fig.\ref{fig:vel_delta0}, where the shock is transferred to a series of compression waves, lies at a certain height from the wall. Therefore the wall pressure gradient is hardly approaching infinity. Combining Eq.\ref{eq:cf_dp_delta0}, $|C_{f,min}| \to \infty$ and singularity in the reversed-flow region is less likely to exist in CCR flows.
	
	\subsection{Peak pressure rise and heat flux}
	
	As discussed above, separation hysteresis results in quite different pressure and heat behavior. Herein, we further discuss the peak pressure rise and peak heat flux in the hysteresis process. 
	
\begin{figure*}[htbp]
		\subfigure[\label{subfig:M6T15P18_S}]{\includegraphics[width = 0.92\columnwidth]{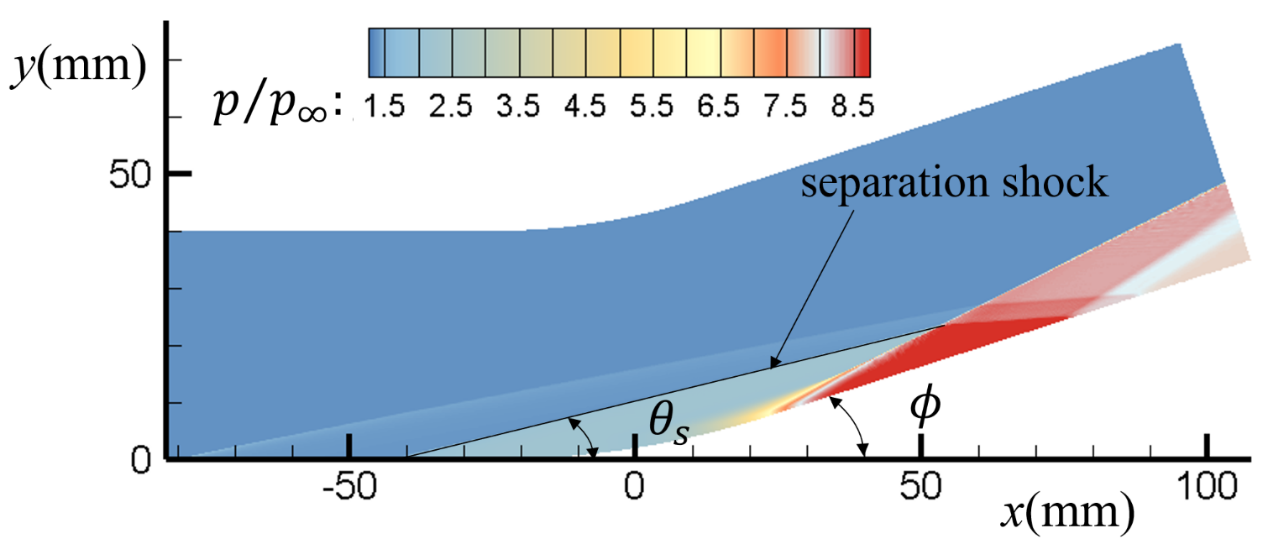}} %\hspace{-2mm}	
		\hspace{8mm}
		\subfigure[\label{subfig:M6T15P18_A}]{\includegraphics[width = 0.92\columnwidth]{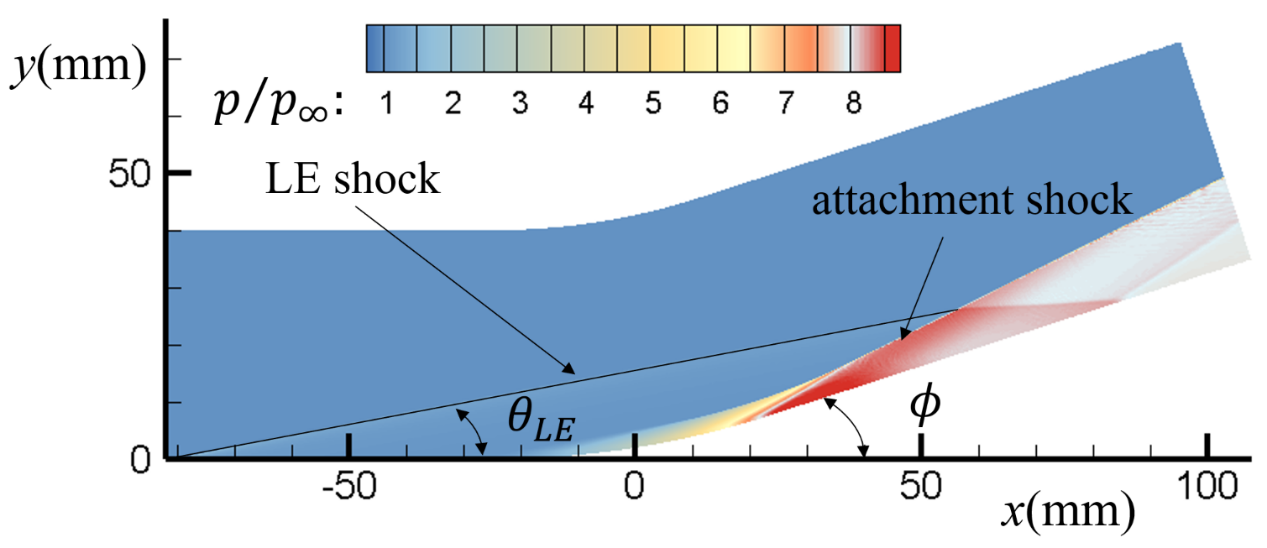}} %\hspace{-2mm}
		
		\subfigure[\label{subfig:M6T15P18_shock_polars}]{\includegraphics[width = 1.9\columnwidth]{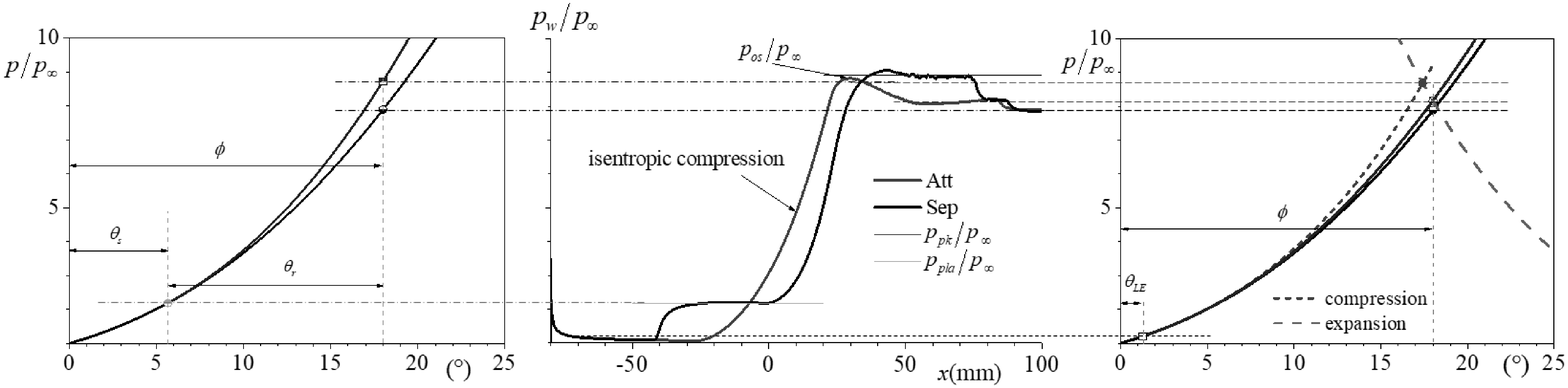}} %\hspace{-2mm}
		\caption{Comparison of DNS results with shock-polar results for separation and attachment states ($Ma_{\infty}=6.0,T_{w}=1.5,\phi=18^\circ$). (a)pressure field of separation state. (b)pressure field of attachment state. (c)shock-polars and DNS results.} \label{fig:shock_polars}
	\end{figure*}

\begin{figure}[htbp]
		\includegraphics[width = 0.9\columnwidth]{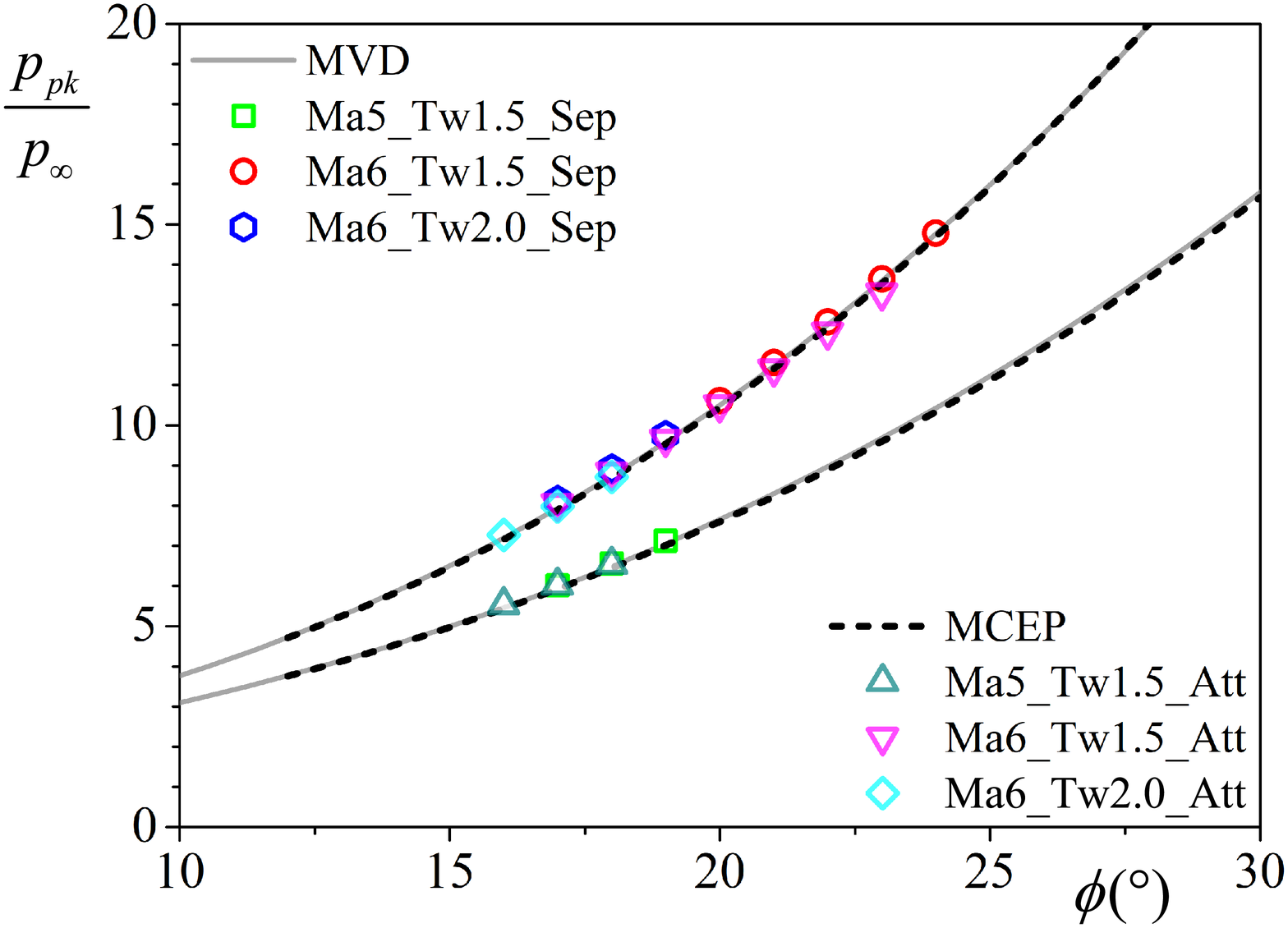} %\hspace{-2mm}	
		%\hspace{4mm}
		\caption{Comparison of results by MVD and MCEP with DNS results.} \label{fig:pmax-MVD3}
	\end{figure}

\begin{figure}[htbp]
		\includegraphics[width = 0.75\columnwidth]{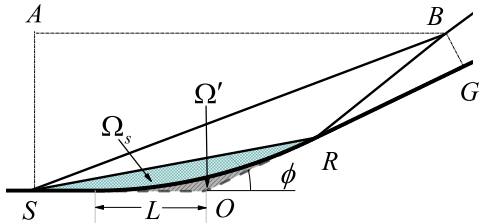} %\hspace{-2mm}	
		\caption{Shock configuration of curved compression ramp flow with large separation.} \label{fig:shock configuration}
	\end{figure}

	For the attachment states, the characteristic pressure can be analyzed by shock-polars and Prandtl-Meyer relation (Fig.\ref{fig:shock_polars}). It is noted that knowledge of the deflection angle $\theta_{LE}$ (Fig.\ref{subfig:M6T15P18_A}) that induces the LE shock is required. And $\theta_{LE}$ is obtained by the post LE shock pressure rise in the present analysis. The pressure presents an overshoot $p_{os}$, which is a result of the pressure match between isentropic compression on the curved wall and the pressure rise by the attachment shock. The pressure match process is implemented by the expansion waves which originate from the reflection of the compression waves on the attachment shock. Good agreement of $p_{os}$ between DNS results and the results based on the match of compression and expansion process (MCEP) is obtained, as shown in Fig.\ref{subfig:M6T15P18_shock_polars} and Fig.\ref{fig:pmax-MVD3} (denoted as $p_{pk} / p_{\infty}$). The post-shock pressure rise (the pressure after attachement shock in Fig.\ref{subfig:M6T15P18_A}) and inviscid pressure rise calculated by shock-polars are also consistent with the DNS results.\\
	Using minimum viscous dissipation (MVD) theorem, Hu et al.\cite{hu2020prediction} have predicted the peak pressure $p_{pk}$ in CRFLS and obtained good agreement with the numerical and experimental results. As the separation states of CCR have similar shock structure to CRFLS’s (Fig.\ref{fig:shock configuration}), the total dissipation $\mathcal{D}$ in the control volume $V$ that consists of SA, AB, BG and the wall is still primarily contributed by separation and reattachment shocks $SB$ and $RB$, i.e.
	\begin{equation} \label{eq:dissp}
		\mathcal{D} = \int_{V} \Phi dV \approx \hat{\Phi}_{SB} L_{SB} + \hat{\Phi}_{RB} L_{RB}
	\end{equation}
	where $L_{SB}$ and $L_{RB}$ are the lengths of separation/reattachment shocks $SB$ and $RB$, respectively. $\hat{\Phi}$ denotes the dissipation induced by a shock per unit length. Furthermore, the assumption in Ref.\cite{hu2020prediction} that the fluid mass in the separation bubble $\Pi_{s} = \rho_{s} \Omega_{s}$ is proportional to the pressure rise $p_{1}$ in the plateau region is still adopted. As $\rho_{s} =\gamma Ma_{\infty}^2 p_{1} / T_{w}$  is proportional to $p_{1}$, the same inference can be drawn that the area of the separation bubble $\Omega_{s}$ is assumed to be constant for given  $Ma_{\infty}$ and $T_{w}$. Besides, as $\Omega^{'} = \frac{L^{2}}{\tan^{2} \frac{\phi}{2}}(\tan \frac{\phi}{2} - \frac{\phi}{2})$ is constant for given $\phi$, the area of triangle $SOR$ $\Omega_{SOR} = \Omega_{s} + \Omega^{'}$ is derived to be constant. Therefore the corresponding model in Ref.\cite{hu2020prediction} is applicable to the CCR flow with large separation.	
	
	The comparison between $p_{pk}/p_{\infty}$ predicted by MVD and the results of DNS is shown in Figs.\ref{subfig:M6T15P18_shock_polars} and Fig.\ref{fig:pmax-MVD3}. Here the deflection angle $\theta_{s}$ in Fig.\ref{subfig:M6T15P18_S} and Fig.\ref{subfig:M6T15P18_shock_polars} that induces the separation shock is also predicted by MVD. We can see that the predicted results are in good agreement with the DNS results for the IBCs selected in this paper. Moreover, $T_{w}$ hardly influence $p_{pk}/p_{\infty}$, indicating that the change of $T_{w}$ has little influence on shock structures in a certain range. The good predictions of peak pressure provide a good basis for predicting peak heat flux in attachment and separation states.
	\par Additionally, an interesting phenomenon can be found in Fig.\ref{fig:pmax-MVD3}. The results of MVD and MCEP overlap well, which means nearly the same peak pressure can be obtained for separation state and attachment state with the same $\phi$ and IBCs. Such phenomenon may indicate the separation bubble shape and the resultant shock structure of separation states result in the minimum visous dissipation to nearly isentropic compression process in the attachment states.
	\par
\begin{figure}[htbp]
		\includegraphics[width = 0.9\columnwidth]{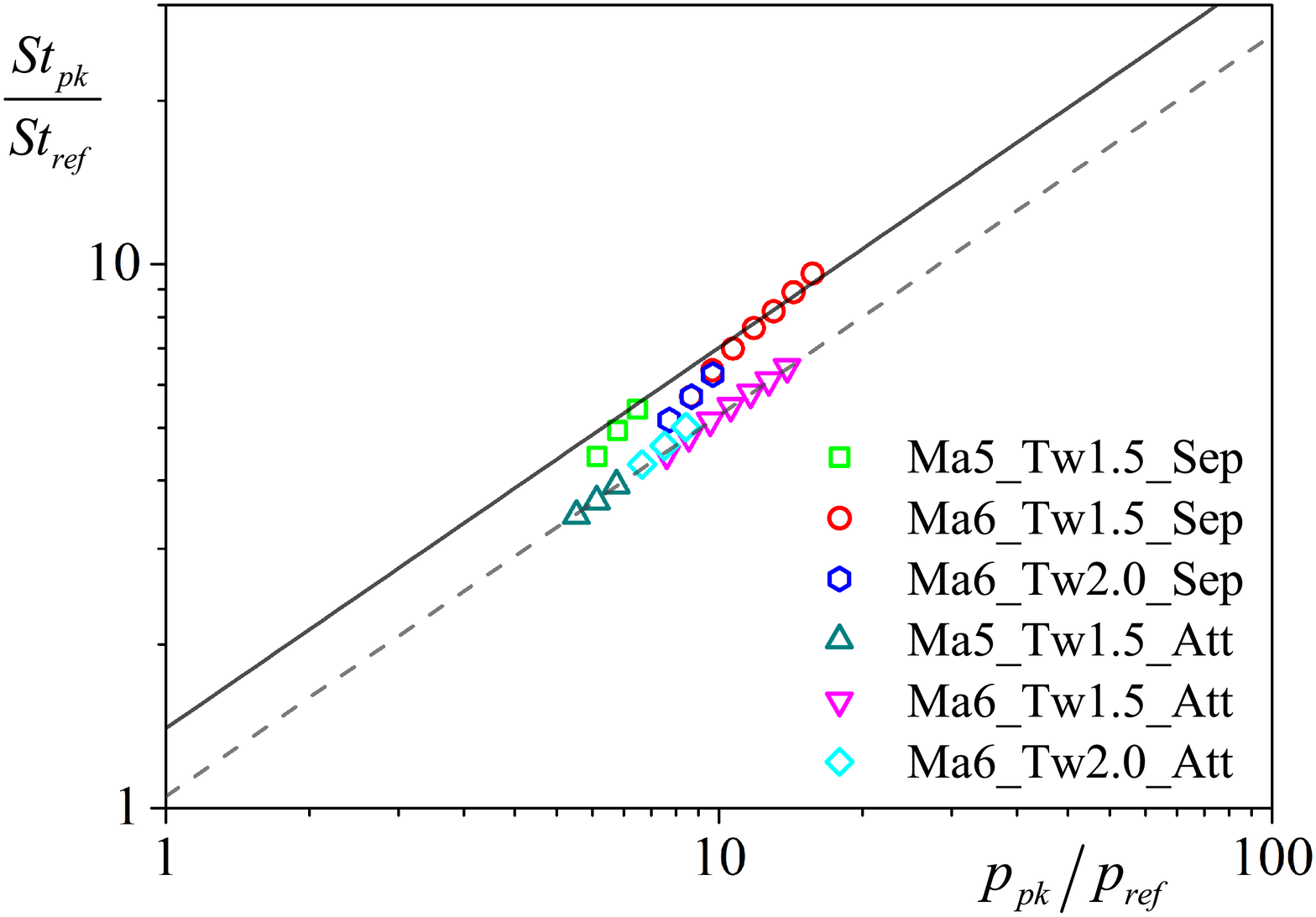} %\hspace{-2mm}	
		\caption{Comparison of simple correlation with DNS results.} \label{fig:stanton-pmax}
	\end{figure}

 Thermal protection design requires the severe peak heat flux in the reattachment region of separation states. There are several methods for predicting the peak heat flux\cite{simeonides1994experimental,simeonides1995experimental,holden1978a,marini2001analysis,2012THEORETICAL,currao2020hypersonic,hung1973shockwave,hung1973interference}, and the peak heat flux is generally correlated with peak pressure rise in these methods. One simple correlation is expressed in the following form\cite{hung1973shockwave,hung1973interference}
	\begin{equation} \label{eq:qpk_corr}
		\frac{St_{pk}}{St_{ref}} = a(\frac{p_{pk}}{p_{ref}})^{b}
	\end{equation}		
	where
	\begin{subequations}\label{eq:qref}
	\begin{align} 
		p_{ref} = &p_{\infty} \\
		St_{ref} = 0.332 &\sqrt{\frac{C^*}{Re_{x_{pk}}}} Pr^{-2/3}
	\end{align}
	\end{subequations}
	The $*$ refers to Eckert's reference temperature method\cite{1960SURVEY}, and $x_{pk}$ is the location of the peak heat flux.
	As shown in Fig.\ref{fig:stanton-pmax}, the separation and attachment states both satisfy Eq.\ref{eq:qpk_corr}, and the exponent $b_{Sep} \approx b_{Att} \approx 0.7$ are indicated by the tendency. The exponent is in accordance with the laminar/laminar interference of Hung\cite{hung1973interference}, while the coefficient $a_{Sep}=1.4>a_{Att}=1.05>1.0$. The discrepancy of coefficient may result from the lower estimate of $St_{ref}$ by Eckert's method\cite{1960SURVEY} than measured values, which can be found in Ref.\cite{hung1973interference}. Larger coefficient of separation states indicates higher peak heat fluxes than that of the attachment states with the same pressure rise. This indicates that it is promising to reduce a mass of heat flux by adjusting $\phi$ or other methods to keep the flow attached in the DSI.

	\section{Conclusion}
	In this paper, separation hysteresis in CCR flow induced by variation of turning angle $\phi$, as well as the corresponding aerothermal characteristics, is investigated. In the interval ($\phi_{a},\phi_{s}$), the boundary layer in CCR flow can be stably in either attachment or separation state with the same IBCs, and the state is determined by the evolutionary history of $\phi$-variation. This interval is the dual-solution interval (DSI). And DSI is shifted and narrowed with the decrease of $Ma_{\infty}$ or the increase of $T_{w}$. For the turning angle outside DSI, only one stable state exists, either attachment or separation. In DSI, the separation and attachment states lead to a great difference in aerothermal features. The pressure rise process in the attachment state is approximately isentropic process, and the peak heat flux from separation state to attachment state can be reduced by up to 32\% within the present cases.
	\par The relationship between the second minimum shear stress $\tau_{2,min}$ and the scaled angle $\alpha$ of the separation states is analyzed. The results indicate that with the increase of $\alpha$, $\tau_{2,min}$ is less likely to reach infinite value for low wall temperature ratio, confirming Gai's suggestion\cite{gai2019hypersonic}. Besides, the emergence of $\tau_{1,min}$ and $\tau_{2,min}$ is interpreted with simplified momentum equation, correlated with APG and characteristic thickness between the wall and the zero shear stress line. The correlation is validated with good agreement, indicating $\tau_{1,min}$ and $\tau_{2,min}$ are dominated by APG induced by separation and reattachment shocks. Results and theoretical analysis indicate that the singularity in the reverse-flow singularities pointed out by Smith\cite{smith1988a} is less likely to appear in CCR flow.
	\par Peak pressure and peak heat flux emerged in the process of separation hysteresis are analyzed. Good prediction of the pressure peak of separation states is obtained with MVD\cite{hu2020prediction}, since the shock structure of CCR separation state is similar with CRFLS. And the pressure rise by separation and reattachment shocks in the separation states is close to that of attachment states by isentropic process for the same $\phi$, indicating the separation bubble shape and the resultant shock structure indeed results in the minimum viscous dissipation. The relationship between peak heat flux and peak pressure rise fulfills the classical power relations, but the peak heat flux in the separation state is significantly higher than that in the attachment state, indicating that maintaining the attachment state in DSI by means of $\phi$ adjustment is of great significance to reduce the heat flux.	
		
		\begin{acknowledgments}
			We are grateful to professor Xin-Liang Li for his helpful discussions. This work was supported by the National Key R \& D Program of China (Grant No. 2019YFA0405300). Yan-Chao Hu thanks to the support of the China Postdoctoral Science Foundation (Grant No. 2020M683746).
		\end{acknowledgments}			
		\section*{reference}
		\nocite{*}
		\bibliography{CCR_sep_hys}% Produces the bibliography via BibTeX.
		%\end{reference}

	\end{document}